\documentclass[lettersize,journal]{IEEEtran}
\usepackage{amsmath,amsfonts,amssymb}
\usepackage{algorithmic}
\usepackage{algorithm}
\usepackage{array}
\usepackage[caption=false,font=normalsize,labelfont=sf,textfont=sf]{subfig}
\usepackage{textcomp}
\usepackage{stfloats}
\usepackage{url}
\usepackage{bbding}
\usepackage{fontawesome}
\usepackage{verbatim}
\usepackage{graphicx}
\usepackage{subcaption}
\usepackage{colortbl}
\usepackage{booktabs}
\usepackage{subcaption}
\usepackage{cite}
\usepackage{color}
\usepackage[table,xcdraw]{xcolor}
\usepackage{multirow}
\usepackage{pifont} 
\usepackage{adjustbox}    % for full-page width centering
\usepackage{datetime}
\newdateformat{myshortdate}{\shortmonthname[\THEMONTH] \THEDAY, \THEYEAR}

\renewcommand{\vec}[1]{\boldsymbol{\mathrm{#1}}}

\newcommand\E{\mathbb{E}}

\begin{document}

%\title{Keep Separate, Train Together: Nonlinear Fusion of Speaker and Spoof Detection for Spoofing-Robust Automatic Speaker Verification}

%\title{Keep Separate, but Optimize Jointly: SASV as Nonlinear Fusion of Speaker and Spoof Detectors}

%\title{Keep Separate, Train Together: %SASV as 
%Nonlinear Fusion of Speaker and Spoof Detectors}

%\title{Unified SASV: Embedding-Based Verification with Learnable Weighted Cosine and MLP Backends}

\title{Joint Optimization of Speaker and Spoof Detectors for Spoofing-Robust Automatic Speaker Verification}

%\title{Integrating Embedding and Score Fusion Techniques for Enhanced Spoofing-Aware Speaker Verification}
%\title{A Hybrid Fusion Approach: Combining Embedding and Score Fusion for Robust SASV}
%\title{A Joint Embedding and Score Fusion Framework for Spoofing-Aware Speaker Verification}
% \title{An Improved Parallel Embedding Fusion with Score-Level Integration for Spoofing-Aware Speaker Verification}

\author{Oğuzhan Kurnaz, Jagabandhu Mishra, Tomi H. Kinnunen, Cemal Hanilçi
        % <-this % stops a space
\thanks{O. Kurnaz (Corresponding author, oguzhan.kurnaz@btu.edu.tr) and C. Hanilçi (cemal.hanilci@btu.edu.tr) are with the Bursa Technical University, Bursa, Turkey.\\ 
J. Mishra (jagabandhu.mishra@uef.fi) and T.H. Kinnunen (tomi.kinnunen@uef.fi) are with the University of Eastern Finland, Joensuu, Finland. \\
This study has been partially supported by the Academy of Finland (Decision No. 349605, project "SPEECHFAKES").}

}

\markboth{Journal of \LaTeX\ Class Files,~Vol.~14, No.~8, August~2021}%
{Shell \MakeLowercase{\textit{et al.}}: A Sample Article Using IEEEtran.cls for IEEE Journals}


\maketitle

\begin{abstract}
Spoofing-robust speaker verification (SASV) combines the tasks of speaker and spoof detection to authenticate speakers under adversarial settings. Many SASV systems rely on fusion of speaker and spoof cues based on independently trained subsystems, which often limits joint performance. In this study, we propose a novel modular, yet jointly optimized, SASV framework that integrates the outputs of speaker and spoofing detection subsystems using trainable back-end classifiers. Our framework enables direct optimization of both subsystems under a unified objective, using the recently-proposed architecture-agnostic detection cost function (a-DCF) as the training objective. This approach preserves the interpretability and plug-and-play compatibility of standalone detectors while aligning them towards a common goal. Our experiments on the ASVspoof 5 dataset demonstrate two important findings: (i) nonlinear score fusion consistently improves a-DCF over linear fusion, and (ii) the combination of weighted cosine scoring for speaker detection with SSL-AASIST for spoof detection achieves state-of-the-art performance, reducing min a-DCF to $0.196$ and SPF-EER to $7.6\%$. These contributions highlight the importance of modular design, calibrated integration, and task-aligned optimization for advancing robust and interpretable SASV systems.
\end{abstract}


\section{Introduction}

Automatic speaker verification (ASV) \cite{reynolds94_asriv} systems are widely deployed in security-critical domains, including banking and call center applications, to authenticate users based on their 
voice characteristics. 
With the successful adoption of advanced deep neural network models, modern ASV systems have become highly effective at distinguishing genuine users (target speakers) from impostors (nontarget speakers). However, they remain vulnerable to spoofing attacks, such as replay \cite{muller25_interspeech}, text-to-speech synthesis (TTS), voice conversion (VC) and adversarial attacks \cite{das20c_interspeech}, all known to compromise system integrity \cite{todisco2019asvspoof}. To counter such threats, various specialized countermeasures (CMs) have been developed for detecting and rejecting artificially generated or manipulated speech. Yet, CM systems alone address only spoofing detection and do not verify speaker identity. 

To address this problem, speaker and spoofing detection can be combined into a unified solution, often referred to as \textbf{spoofing-robust automatic speaker verification} (SASV). Early studies in the mid-2010s explored joint use of ASV and spoofing countermeasures (e.g., with i-vector speaker embeddings \cite{sizov2015}). Since 2019, the ASVspoof challenge series has also addressed SASV task through a specific, cascaded ASV+CM architectures with a fixed ASV system, using a tailored performance metric \cite{tdcf_kinnunen2020}.Later on, the SASV2022 challenge \cite{jung2022sasv} promoted development of SASV architectures not limited to cascaded architectures. Similarly, the ASVspoof 5 challenge \cite{Wang2024_ASVspoof5} also featured a submission track for arbitrary SASV architectures, including more advanced adversaries and broader speaker diversity. 

Methodologically, existing SASV approaches---reviewed in further detail in Section~\ref{sec:existing-sasv-models}---can be grouped into two broad categories of \emph{end-to-end}~\cite{teng22_interspeech, kang22_interspeech} and \emph{modular}~\cite{todisco18_interspeech,Wu2022e} systems. Whereas the former maps a speech signal directly to a SASV score, the latter combines the outputs of independently developed ASV and CM subsystems, either at the embeddings or score level \cite{Shim2022}. While the former benefits from holistic optimization, it is traded for limited interpretability and accountability, as it becomes unclear whether errors arise from speaker or spoof detection failures. Embedding fusion-based systems combine intermediate vector space representations produced by ASV and CM systems and use them as inputs to a back-end classifier. Score fusion, in turn, combines the detection scores or hard binary decisions of the two subsystems. This modular design improves interpretability and flexibility of the resulting SASV system.

\begin{figure}[!t]
  \centering
  \includegraphics[scale=0.24]{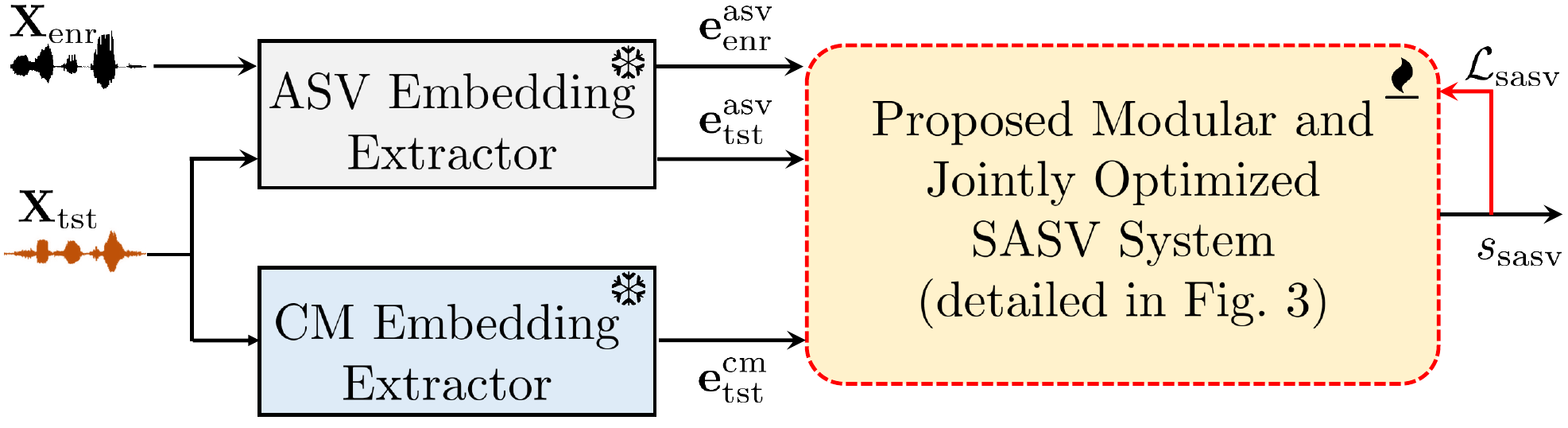}
    \caption{Overview of the proposed modular and jointly optimized SASV framework. Embeddings, extracted from enrollment and test utterances ($\mathbf{X}_\text{enr}$ and $\mathbf{X}_\text{tst}$) using frozen (\SnowflakeChevronBold) ASV and CM extractors, are fed into the modular SASV system (red dashed box). 
    Its internal modules are jointly optimized with the SASV loss $\mathcal{L}_{\mathrm{sasv}}$ (see Fig.~\ref{fig:your_label} for details), producing the final SASV score $s_{\mathrm{sasv}}$. Modules marked with \faFire\ denote trainable components.} 
  \label{fig:proposed_approach}
\end{figure}

It is instructive to contrast fusion approaches used in conventional ASV vs. SASV systems, as there is a subtle but important difference between them. In the former case, the combined subsystems all address the \emph{same} (i.e. speaker detection) task. Fusion of ASV and CM, however, involves \emph{two different} tasks (i.e. speaker and spoof detection). As a result, established popular fusion recipes used in ASV research, such as (weighted) averaging of detector scores, are inapplicable or suboptimal for SASV. A recent work \cite{wang24l_interspeech} showed, both theoretically and experimentally, that post-processing \cite{leeuwen13_interspeech} ASV and CM scores into calibrated log-likelihood ratios (LLRs) prior to fusion improves performance. Commonplace linear fusion strategies struggle to capture the complex non-linear dependencies arising from ASV and CM interaction under adversarial or unseen conditions. Consequently, non-linear fusion approaches \cite{todisco18_interspeech,wang24l_interspeech, rohdin24_asvspoof} overcome this limitation by learning data-driven interactions between scores, enabling more discriminative decision boundaries and improved robustness. 

In this work, we propose a novel modular, yet jointly optimized SASV architecture. Unlike purely end-to-end systems, our method preserves modularity (interpretability and plug-and-play compatibility of standalone speaker and spoof detectors). Advancing upon previous modular designs, however, we address \textbf{direct joint optimization of both subsystems under the SASV objective}, aligning them towards a common goal. The integration is achieved via a non-linear score fusion module operating on calibrated ASV and CM scores, following solid decision theory principles (Section~\ref{sec:decision-theory-background}).

To align optimization with operational requirements, we adopt the architecture-agnostic detection cost function (a-DCF) \cite{kurnaz2024optimizing} as the primary training objective, ensuring that optimization reflects the practical trade-offs between target, nontarget, and spoof trials. Auxiliary binary cross-entropy losses guide specialization of the individual subsystems, balancing global optimization with subsystem reliability. We also systematically investigate three ASV scoring designs—MLP classification, cosine similarity and its learnable weighted variant—to analyze their robustness under different spoofing conditions. Fig.~\ref{fig:proposed_approach} illustrates the general framework for the proposed SASV architecture, with full details provided in Section~\ref{sec:proposed-approach}. 

Compared to our earlier preliminary work in \cite{kurnaz2024optimizing}, which focused on optimizing either embedding concatenation or non-linear score fusion of independently trained ASV (ECAPA-TDNN) and CM (AASIST) systems using the a-DCF objective, the present study extends SASV in two key directions. First, we introduce direct joint optimization of ASV and CM subsystems under a unified SASV objective, rather than training them independently. Second, we systematically investigate alternative ASV scoring designs (MLP, cosine similarity, and weighted cosine similarity) and analyze their effectiveness under joint optimization, thereby generalizing and broadening the scope of our previous approach.

We intend our paper to be self-contained in that it describes the decision-theoretic motivation for SASV (Section~\ref{sec:decision-theory-background}), presents a generic modular framework, and combines direct optimization with extensive experimental validation using state-of-the-art ASV backbones (ECAPA-TDNN, WavLM-TDNN, ReDimNet) and CM models (AASIST, SSL-AASIST) under challenging spoofing scenarios. By jointly addressing subsystem integration, optimization and robustness, the proposed framework offers a practical and effective solution for SASV, capable of generalizing to challenging and previously unseen spoofing scenarios.

\section{Decision-Theoretic Background to SASV}\label{sec:decision-theory-background}

SASV combines the tasks of speaker \cite{reynolds1995} and spoofing \cite{wu2015} detection to facilitate user authentication under scenarios where spoofing is anticipated to take place. Following \cite{wang24l_interspeech}, we cast SASV under principled, decision-theoretic formulation, beginning from the standalone tasks of speaker and spoofing detection.

\subsection{Conventional ASV (speaker detection without spoofing)}

The task of ASV \cite{reynolds1995} is to determine whether a given speech utterance $\vec{X}$ matches a claimed speaker identity (target speaker) or not (non-target speaker). In this binary classification (detection) setting, exactly one of the two exhaustive and mutually exclusive %hypotheses, 
propositions, denoted by 
\begin{equation}
    \mathcal{Y}_\text{ASV} := 
    \left\{
    \begin{array}{ll}
        y_\text{tar} & : \text{target speaker present}, \\
        y_\text{non} & : \text{non-target speaker present}
    \end{array}
    \right\},
\end{equation}
is assumed true\footnote{ The 'target' and 'non-target' terminology follows established nomenclature used in ASV literature, being synonymous for 'same speaker' and 'different speaker', respectively. This speaker similarity flag is defined independent of whether $\mathbf{X}$ originates from a bonafide human or a spoofing system. An example of 'targeted' spoofing attacks is voice conversion.}. By viewing an ASV system %is viewed as a 
as a rational decision making agent, our task is to \emph{choose an optimal action} $a \in \mathcal{A}$ from the set of allowed actions $\mathcal{A}$ (i.e. make a decision). In many\footnote{One exception are ASV systems that conclude $\vec{X}$ to be too noisy or too short (or otherwise unreliable) for making a reliable decision. In this case it is natural to include a third action 'no decision'. This example contains three possible actions but the number of classes remains two; the number of classes and decisions do not have to match.} cases, including this work, the actions are either to accept or reject the identity claim. We denote the binary action set by 
$\mathcal{A} := \{\text{accept},\text{reject}\}$.

The aim in statistical decision theory~\cite{duda2000pattern,Jaynes03-prob-theory} is to choose an %optimal 
action $a_* \in \mathcal{A}$ such that 
    \begin{equation}\label{eq:minimum-risk-decision}
        \begin{aligned}
            a_* & = \arg\min_{a \in \mathcal{A}} R(a|\vec{X})\\
            R(a|\vec{X}) & := \E_{P(y|\vec{X})}\left[ C(a,y)\right] = \sum_{y} C(a,y)P(y|\vec{X}),  
        \end{aligned}
    \end{equation}
where $R(a|\vec{X})$ is \emph{conditional risk} for taking an action $a$ for a given input $\vec{X}$, $y \in \mathcal{Y}$ 
is class label, $P(y|\vec{X})$ is class posterior and $\E_{P}\left[ \cdot \right]$ is the expected value with respect to probability distribution $P$. Finally, $C: \mathcal{A}\times \mathcal{Y} \rightarrow \mathbb{R}^+$ is a nonnegative \emph{decision cost function} that assigns value $C(a,y)$ for taking an action $a$ when the actual class is $y$. 

With the two-class and two-action ASV task concerned herein, $C(a,y)$ is represented by a $2 \times 2$ matrix where the actions and classes are organized on rows and columns, respectively. Correct decisions (diagonal entries) are assigned a cost of 0. The two remaining cases correspond to acceptance of a non-target speaker (\emph{false acceptance} or \emph{false alarm}) and rejection of the target speaker (\emph{false rejection} or \emph{miss}). The costs for these error cases are arbitrary constants that reflect the desired error trade-off behavior and which %the decision making agent 
one must fix in advance. By adapting shorthands $C_\text{fa}^\text{non} \equiv C(\text{accept},\mathcal{H}_\text{non})$ and $C_\text{miss}^\text{tar} \equiv C(\text{reject},\mathcal{H}_\text{tar})$, the conditional risks for the two actions are written as
    \begin{equation}
        \begin{aligned}
            R(\text{accept}|\vec{X}) & = C_\text{fa}^\text{non}P(y_\text{non}|\vec{X})\\
            R(\text{reject}|\vec{X}) & = C_\text{miss}^\text{tar}P(y_\text{tar}|\vec{X}).
        \end{aligned}
    \end{equation}
By following the minimum-risk strategy in \eqref{eq:minimum-risk-decision} and applying Bayes rule, it is easy to show that the optimal decision policy is to accept the speaker if and only if $\ell_\text{non}^\text{tar}(\vec{X}) > \tau_{\text{ASV}}^{\text{Bayes}}$ \cite[Sec.~2]{duda2000pattern}, where 
    \begin{equation}\label{eq:optimal-decision-ASV}
        \begin{aligned}
            \ell_\text{non}^\text{tar}(\vec{X}) & := \log \frac{p(\vec{X}|y_\text{tar})} {p(\vec{X}|y_\text{non})}\\
            \tau_{\text{ASV}}^{\text{Bayes}} & := \log(C_\text{fa}^\text{non}/C_\text{miss}^\text{tar})-\text{logit}(\pi_\text{tar})\
        \end{aligned}
    \end{equation} 
denote, respectively, the target-to-nontarget \emph{log-likelihood ratio} (LLR) score and the Bayes decision threshold. Here, $\pi_{\text{tar}} = P(y_{\text{tar}})$ is the prior probability of the target speaker being present and $\text{logit}(\pi) := \log(\pi) - \log(1 - \pi)$. Note that whereas the LLR score depends on the observed data $\vec{X}$, the decision threshold is solely determined from the decision costs and class priors. As an example, for equally costly error types ($C_\text{fa}^\text{non}=C_\text{miss}^\text{tar}$) and flat prior ($\pi_\text{tar}=0)$, we have $\tau_\text{ASV}^\text{Bayes}=0$, i.e. the decision rule is to accept the speaker if (and only if) the LLR score is positive. 

While Bayes' decision theory provides a normative framework for making rational, statistically optimal decisions, the optimality relies on knowledge of the true probability distributions. This is generally unreasonable. Further, not all classifiers produce LLRs or have even a direct probabilistic interpretation---familiar example being cosine similarity between a pair of enrollment and test speaker embeddings. Widely studied in ASV \cite{Brummer2010}, particularly in its forensic applications \cite{RAMOS2013-reliable-support,Morrison2013-calibration}, the practical remedy is to \emph{calibrate} arbitrary speaker similarity scores $s_\text{asv}(\vec{X})$ as a post-processing operation, so that the calibrated scores can be effectively treated as (calibrated) LLRs \cite{brummer13_interspeech, leeuwen13_interspeech, ferrer20b_odyssey}. There are many approaches for score calibration, a common approach being an affine transform $w_0 + w_1 s_\text{asv}(\vec{X})$, where the parameters $w_0$ and $w_1$ are trained using labeled training trials.

\subsection{Spoofing detection}

Spoofing detection \cite{wu2015} aims to determine whether an audio input is bonafide (real) or spoofed (fake). Even if the features and detection models are usually different from ASV, the optimal decision making strategy outlined above remains applicable. For our purposes, the only relevant difference is in the class labels, which are now
\begin{equation}
    \mathcal{Y}_\text{CM} := 
    \left\{
    \begin{array}{ll}
        y_\text{bon} & : \text{input is bonafide (real) speech}, \\
        y_\text{spf} & : \text{input is spoofed (fake) speech}
    \end{array}
    \right\}.
\end{equation}
 
An ideal CM should accept all bonafide utterances and reject all spoofed utterances. A practical CM system takes a speech utterance $\vec{X}$ as input and outputs a score \( s_{\text{cm}}(\vec{X}) \) which reflects 'realness' of the input utterance, and which is subsequently compared against a decision threshold. Again, the CM score may (or may not) have an interpretation as an LLR score, with standard calibration methods being applicable. 

\subsection{Spoofing-robust speaker verification (SASV)}

Whereas the above decision making strategy for binary classification is well-known, optimal decisions for SASV appear somewhat less known to the community. In fact, the aim of SASV is no different from ASV: to accept or reject an identity claim based on evidence $\vec{X}$. In contrast to conventional ASV, however, it is acknowledged that spoofing attacks may be presented to the system. Spoofed utterances are considered to be outside of the
normal presentation mode of biometric verification~\cite{ISOpresentationAtack}. Formally, SASV is a three-class task where the spoofing attacks form the added class on top of target and non-target classes. 

Cartesian product of the two label sets $\mathcal{Y}_\text{ASV} \times \mathcal{Y}_\text{CM}$ leads to four possible cases that an SASV system may encounter. %Following prior studies, 
The three classes of interest in authentication scenarios are
\begin{itemize}
    \item $y_{\text{tar.bon}} := y_{\text{tar}} \land y_{\text{bon}}$, bonafide target speaker
    \item $y_{\text{non.bon}} := y_{\text{non}} \land y_{\text{bon}}$, bonafide non-target speaker
    \item $y_{\text{spf}}$, spoofed utterance (whether target or non-target),
\end{itemize} 
where $\land$ denotes the logical AND operator. When spoofing is not present ($y_\text{bon}$ is identically true) the ground-truth labeling reduces to the two conventional target and non-target labels.

\begin{table}[h]
\centering
\caption{Values of decision cost function $C(a,y)$ for SASV.}
\begin{tabular}{@{}r@{\hspace{1em}}cc@{}}
\toprule
\textbf{True class label $y$} & \multicolumn{2}{c}{\textbf{Action $a$}} \\
 & \textit{accept} & \textit{reject} \\
\midrule
tar.bon & 0 & $C_{\text{miss}}^{\text{tar.bon}}$ \\
non.bon & $C_{\text{fa}}^{\text{non.bon}}$ & 0 \\
spf    & $C_{\text{fa}}^{\text{spf}}$ & 0 \\
\bottomrule
\label{tab:cost_matrix} 
\end{tabular}
\end{table}

As with conventional ASV, an SASV system should select an action $a \in \{\text{accept}, \text{reject}\}$ that minimizes the conditional risk in \eqref{eq:minimum-risk-decision}. With the added class of spoofing attacks, the matrix that represents the decisions costs now has 6 values (3 classes $\times$ 2 actions). Following \cite{wang24l_interspeech}, using the decision cost notations shown in Table \ref{tab:cost_matrix}, the conditional risks in SASV are now
\begin{align*}
R(\text{accept} \mid \vec{X}) &= C_{\text{fa}}^{\text{non.bon}} P(y_{\text{non.bon}} \mid \vec{X})
 + C_{\text{fa}}^{\text{spf}} P(y_{\text{spf}} \mid \vec{X}) \\
R(\text{reject} \mid \vec{X}) &= C_{\text{miss}}^{\text{tar.bon}} P(y_{\text{tar.bon}} \mid \vec{X})
\end{align*}
where $P(\cdot|\vec{X})$ are the class posteriors. The three costs $C_{\text{fa}}^{\text{non.bon}}$,  $C_{\text{fa}}^{\text{spf}}$ and $C_{\text{miss}}^{\text{tar.bon}}$ denote the costs of falsely accepting bonafide non-target, falsely accepting spoofed utterance, and falsely rejecting bonafide target speaker, respectively. Using Bayes' theorem, the condition for identity claim acceptance becomes \begin{align}\label{eq:sasv-decision-equation}
C_{\text{miss}}^{\text{tar.bon}} \, p(\mathbf{X} \mid y_{\text{tar.bon}})\, \pi_{\text{tar.bon}} 
&> C_{\text{fa}}^{\text{non.bon}} \, p(\mathbf{X} \mid y_{\text{non.bon}})\, \pi_{\text{non.bon}} \nonumber \\
&\quad + C_{\text{fa}}^{\text{spf}} \, p(\mathbf{X} \mid y_{\text{spf}})\, \pi_{\text{spf}},
\end{align}
where the $\pi_{\bullet}$ are the priors of the three classes. To obtain an expression in terms of ASV and CM likelihood ratios, let us rewrite \eqref{eq:sasv-decision-equation} as
\begin{equation}\label{eq:optimal-SASV-decision}
\begin{split}
\pi_{\text{tar.bon}} > &
\;\frac{C_{\text{fa}}^{\text{non.bon}} \, p(\mathbf{X} \mid y_{\text{non.bon}})}
        {C_{\text{miss}}^{\text{tar.bon}} \, p(\mathbf{X} \mid y_{\text{tar.bon}})} \, \pi_{\text{non.bon}} \\
&+ \frac{C_{\text{fa}}^{\text{spf}} \, p(\mathbf{X} \mid y_{\text{spf}})}
        {C_{\text{miss}}^{\text{tar.bon}} \, p(\mathbf{X} \mid y_{\text{tar.bon}})} \, \pi_{\text{spf}} .
\end{split}
\end{equation}

To rewrite this decision rule in terms of LLRs, let 
\begin{align}
\ell^{\text{tar.bon}}_{\text{non.bon}}(\vec{X}) & := \log \frac{p(\vec{X} \mid y_{\text{tar.bon}})}{p(\vec{X} \mid y_{\text{non.bon}})} \label{eq:asv_llr} \\
\ell^{\text{tar.bon}}_{\text{spf}}(\vec{X}) & := \log \frac{p(\vec{X} \mid y_{\text{tar.bon}})}{p(\vec{X} \mid y_{\text{spf}})},
\label{eq:cm_llr}
\end{align}
denote the LLRs for the standalone ASV and CM tasks, respectively. Additionally, let
\begin{equation}
\rho := \frac{\pi_{\text{spf}}}{\pi_{\text{non.bon}} + \pi_{\text{spf}}}
\label{eq:rho}
\end{equation}
denote \emph{spoof prevalence prior}~\cite[Section 3.3]{Kinnunen2024-tEER}---relative proportion of spoofing attacks within the combined class of non-target speakers and spoofing attacks. Finally, define
\begin{equation}
\beta := \frac{\pi_{\text{tar.bon}}}{1 - \pi_{\text{tar.bon}}},
\label{eq:beta}
\end{equation}
so that $\rho$ and $\beta$ collectively re-parameterize the prior distribution $\vec{\pi}=(\pi_\text{tar.bon},\pi_\text{non.bon},\pi_\text{spf})$. With these definitions, a decision rule equivalent to \eqref{eq:optimal-SASV-decision} can be written as follows:
\begin{center}
    \textbf{Optimal SASV Decision Policy}
\end{center}
\begin{equation}
\footnotesize
\label{eq:optimal-SASV-decision-rewritten}
\boxed{
-\log
\left[
 (1 - \rho)\frac{C_{\text{fa}}^{\text{non.bon}}}{C_{\text{miss}}^{\text{tar.bon}}}
e^{-\ell^{\text{tar.bon}}_{\text{non.bon}}(\vec{X})} +
 \rho \frac{C_{\text{fa}}^{\text{spf}}}{C_{\text{miss}}^{\text{tar.bon}}}
e^{-\ell^{\text{tar.bon}}_{\text{spf}}(\vec{X})}
\right]
> -\log \beta}
%\label{eq:flipped_score}
\end{equation}

Setting the spoof prior $\rho$ to its extreme values provides useful insight. When $\rho = 0$ (no spoofing assumed), the SASV decision rule reduces to the standard target--nontarget LLR. In contrast, when $\rho = 1$, the decision depends only on the spoof--target LLR, focusing on distinguishing spoofed trials from target bonafide speech. This corresponds to the spoof-detection component of SASV. Under the assumption that both target and non-target speakers (both of which are bonafide) produce identical bonafide-spoof LLR distributions, this corresponds to a standalone CM system.

There is one important difference to the optimal decision making in conventional ASV. Whereas in  \eqref{eq:optimal-decision-ASV} all the data-related terms (i.e. the LLR score) appear on one side of the inequality and all decision policy related terms (i.e. the threshold) on the other side, this is not the case for \eqref{eq:optimal-SASV-decision-rewritten}: the left-hand side contains expressions that depend both on $\vec{X}$ and the cost model parameters. As discussed in~\cite{wang24l_interspeech}, it is not possible to decouple the LLRs in the same way as in conventional ASV. To mitigate this entanglement, score calibration methods such as those used in~\cite{rohdin24_asvspoof} apply logistic regression to better align LLR distributions, enabling more stable thresholding even when the decision function cannot be decoupled cleanly.

\subsection{On Score Fusion of ASV and CM}

On the basis of the left-hand side in \eqref{eq:optimal-SASV-decision-rewritten}, \cite[Eq. (11)]{wang24l_interspeech} defined a non-linear score fusion approach,
    \begin{equation}\label{eq:basic-nonlinear-fusion}
        s_\text{sasv} = -\log\left[(1-\tilde{\rho})e^{-\ell^{\text{tar.bon}}_{\text{non.bon}}(\vec{X})} +
 \tilde{\rho} e^{-\ell^{\text{tar.bon}}_{\text{spf}}(\vec{X})}\right],
    \end{equation}
where $\tilde{\rho}$ is a tunable fusion parameter. Clearly, $s_\text{sasv}$ is monotonically increasing in both %the 'ASV' and 'CM' LLRs---
LLRs---when either %score 
increases, the likelihood of SASV system accepting the identity claim increases. Note that \eqref{eq:basic-nonlinear-fusion} aligns with the left-hand side of \eqref{eq:optimal-SASV-decision-rewritten}, if one chooses $\tilde{\rho}=\rho$ and further sets $C_\text{fa}^\text{non.bon}=C_\text{miss}^\text{tar.bon}$ and $C_\text{fa}^\text{spf}=C_\text{miss}^\text{tar.bon}$. With the further assumption that each of the three class (target bonafide, non-target bonafide, spoof) are modeled as Gaussians, \eqref{eq:basic-nonlinear-fusion} can shown to be equivalent with the so-called \emph{Gaussian back-end fusion} \cite{todisco18_interspeech}. 

\begin{figure}[htbp]
    \centering
    \includegraphics[scale = 0.44]{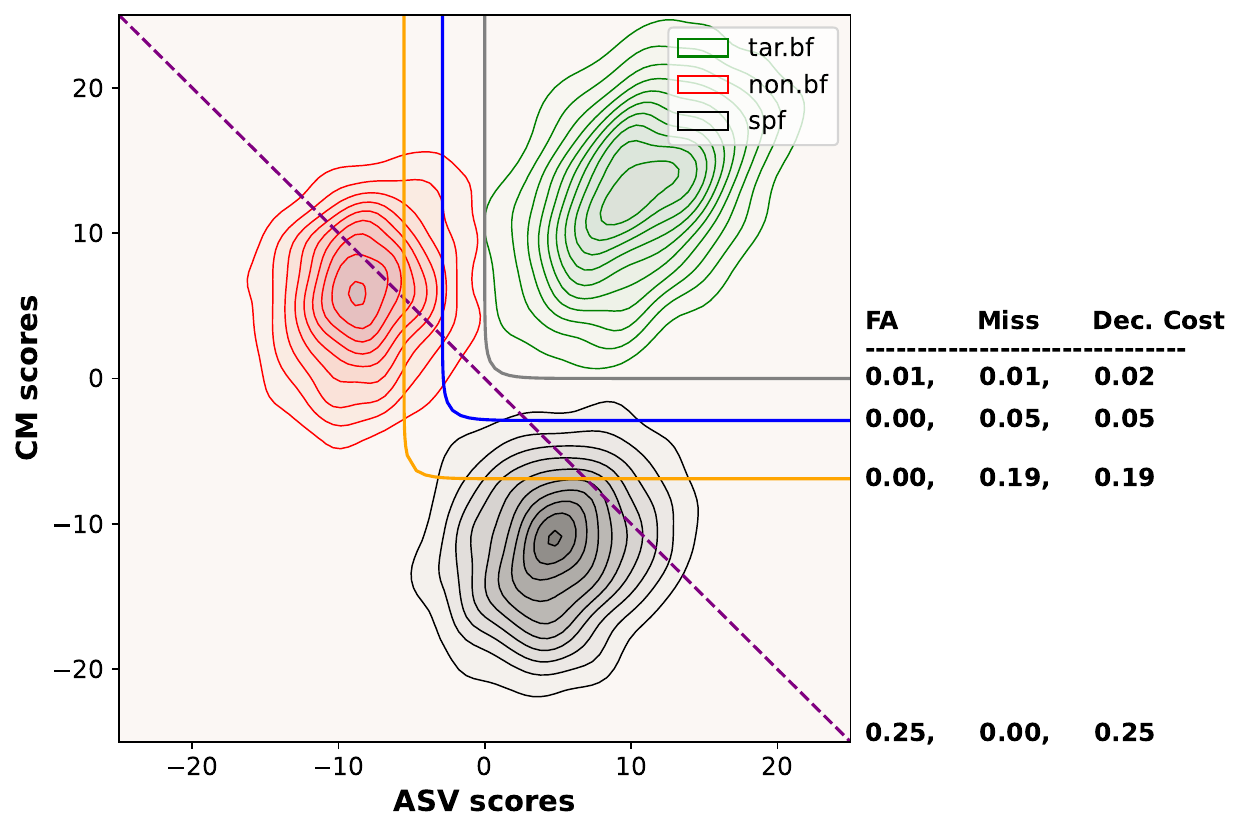}
    \caption{Contour plot of calibrated ASV and CM scores for target, non-target, and spoof trials from simulated data. The dashed \textcolor{purple}{purple} line shows the linear decision boundary assuming uniform priors ($\pi_{\text{tar}} = \pi_{\text{non}} = \pi_{\text{spf}} = \frac{1}{3}$), while the solid \textcolor{gray}{gray} line shows the non-linear boundary from \eqref{eq:basic-nonlinear-fusion} under the same equal-prior assumption. The \textcolor{blue}{blue} and \textcolor{orange}{orange} boundaries correspond to priors $(0.9, 0.05, 0.05)$ and $(0.995, 0.004, 0.001)$, respectively. The numerical values on the right-hand side indicate the error rates for each decision boundary, where ''miss'' denotes the rejection of target trials and ''FA'' (false alarm) denotes the acceptance of non-legitimate trials (non-targets and spoofs). The resulting ``Dec. Cost'' is computed as the sum of these two error terms.}
    \label{fig:lin_nonlin_fusion}
\end{figure}

While \eqref{eq:optimal-SASV-decision-rewritten} provides the optimal decision policy for SASV, simple linear score fusion involving sum or average of raw ASV and CM scores has been more popular in practice. This method assumes equal contribution of ASV and CM, and hence does not take into account potential differences neither in class discrimination nor the numerical scale of the two types of scores. To overcome these limitations, \cite{wang24l_interspeech} proposed a more principled fusion framework based on so-called \emph{compositional data analysis}.  
The insight from~\cite{wang24l_interspeech} is that, rather than summing up raw (uncalibrated) scores, one should average LLRs:
\begin{equation}\label{eq:linear-fusion-of-LLRs}
s_{\text{sasv}} = \frac{1}{\sqrt{6}}
\left(
  \ell^{\text{tar.bon}}_{\text{non.bon}}(\vec{X})
  + \ell^{\text{tar.bon}}_{\text{spf}}(\vec{X})
\right).
\end{equation}

where the constant $1/\sqrt{6}$ originates from the so-called \emph{isometric log ratio} (IRL) transform \cite{Egozcue2003-ILR} and does not impact discrimination. Since arbitrary ASV and CM scores rarely present calibrated LLRs, the original raw scores should be calibrated before averaging. Despite its appeal as an intuitive and simple linear fusion method, \eqref{eq:linear-fusion-of-LLRs} does not yield a Bayes-optimal decision policy for SASV, \emph{even when the true LLRs are known}. Supported further by the experimental comparisons in~\cite{wang24l_interspeech}, linear fusion of LLRs was found inferior to non-linear strategies. As an intuitive geometric example displayed in Figure~\ref{fig:lin_nonlin_fusion}, the linear fusion method does not adequately separate spoofed and non-target samples from the target trials. In contrast, the non-linear fusion approach produces a non-linear decision boundary (blue curve) that leads to improved discrimination of the bonafide targets trials from the two other classes. 

To sum up, both the theoretical and empirical evidence points that the premise for designing modular SASV systems should use \eqref{eq:optimal-SASV-decision-rewritten} (rather than \eqref{eq:linear-fusion-of-LLRs}) as the foundational basis. For both pedagogical and contrastive purposes, we nonetheless compare both types of approaches in our experiments. 

\subsection{a-DCF loss}
\label{sec:adcf_loss}
With the decision-theoretic foundations laid out above, two important practical considerations remain: (1) how to \emph{evaluate} SASV performance; and (2) how to \emph{optimize} an SASV system? The two questions are linked since, ideally, a classifier would be directly optimized for the metric it is being assessed on. However, the standard a-DCF is non-differentiable; therefore, as detailed below, we utilize a differentiable surrogate (the soft a-DCF loss) for optimization, while all performance results reported in Section~\ref{sec:experimental_results} are computed using the original a-DCF formulation.

While numerous standard nonparametric metrics such as accuracy, F1 score and EER are available, none are aligned with optimal decision making. Moreover, these metrics are designed to assess binary classifiers, making them unsuitable for ternary tasks like SASV. For instance, many SASV studies report EER between bonafide targets against the pooled negative class consisting of bonafide non-targets and spoofing attacks. This leads to EER being dependent on empirical nontarget-spoof class proportions ~\cite[Section 4.4]{Kinnunen2024-tEER}; see also ~\cite[Section 2]{shim2024adcf}.

The authors in~\cite{shim2024adcf} proposed \emph{architecture-agnostic detection cost function} (a-DCF) for performance assessment of SASV systems, extending cost-based assessment of conventional ASV systems~\cite{Doddington2000-NIST}. Different both from the conventional DCF~\cite{Doddington2000-NIST}---limited to binary classification---and the `tandem DCF' (t-DCF)~\cite{tdcf_kinnunen2020}---limited to particular cascaded ASV and CM fusion architecture---the a-DCF metric is applicable to SASV with any architecture that outputs a single detection score, $s_\text{sasv}$. The a-DCF measures the expected cost of decisions, formalized as 

\begin{equation}
\label{adcf_old}
\begin{aligned}
\mathrm{a\!-\!DCF}(\tau_\text{sasv}) = & \; C_{\text{miss}}^{\text{tar.bon}} \cdot \pi_{\text{tar}} \cdot P_{\text{miss}}^{\text{tar.bon}}(\tau_\text{sasv}) \\
& + C_{\text{fa}}^{\text{non.bon}} \cdot \pi_{\text{non}} \cdot P_{\text{fa}}^{\text{non.bon}}(\tau_\mathrm{sasv}) \\
& + C_{\text{fa}}^{\text{spf}} \cdot \pi_{\text{spf}} \cdot P_{\text{fa}}^{\text{spf}}(\tau_\text{sasv}),
\end{aligned}
\end{equation}

\noindent where the costs and priors are as in Table~\ref{tab:cost_matrix}. The three error rates $P_{\text{miss}}^{\text{tar.bon}}(\tau_\text{sasv})$, $P_{\mathrm{fa}}^{\text{non.bon}}(\tau_\text{sasv})$ and  $P_{\text{fa}}^{\text{spf}}(\tau_\text{sasv})$ are the bonafide target miss, bonafide non-target false alarm and spoof false alarm rates, respectively. All are functions of a detection threshold $\tau_\text{sasv}$. They are estimated by error counting:
{\small
\begin{equation}
\begin{aligned}
P_{\text{miss}}^{\text{tar.bon}}(\tau_\text{sasv}) &= \frac{1}{N_\text{tar.bon}} \sum_{\vec{X} \in \text{tar}}
H(\tau_\text{sasv} - s_\text{sasv}(\vec{X})) \\
P_{\text{fa}}^{\text{non.bon}}(\tau_\text{sasv}) &= \frac{1}{N_\text{non.bon}} \sum_{\vec{X} \in \text{non}}
H(s_\text{sasv}(\vec{X}) - \tau_\text{sasv}) \\
P_{\text{fa}}^{\text{spf}}(\tau_\text{sasv}) &= \frac{1}{N_\mathrm{spf}} \sum_{\vec{X} \in \mathrm{spf}}
H(s_\text{sasv}(\vec{X}) - \tau_\text{sasv}),
\end{aligned}
\label{eq:all_p}
\end{equation}}
where $s_\text{asv}(\textbf{X})$ is the SASV score for trial $x$, and where $\mathrm{tar}$, $\mathrm{non}$ and $\mathrm{spf}$ denote the sets of target, non-target, and spoof trials, respectively, with their counts denoted by $N_\bullet$. Here, $H(\cdot)$ is the heaviside step function with the $H(t)=0$ for $t < 0$ and $H(t)=1$ for $t \geq 0$, used for error counting.

In our recent work~\cite{kurnaz2024optimizing}, we addressed SASV optimization directly for the a-DCF metric. In practice, the heaviside function $H(\cdot)$ in \eqref{eq:all_p} is replaced by its differentiable approximation, namely, the logistic sigmoid $\sigma(t) = 1/\left(1 + \mathrm{exp}(-t)\right)$, following the approach proposed in \cite{mingote19_interspeech} and adopted in our recent work~\cite{kurnaz2024optimizing}. %\frac{1}{1 + e^{-t}}$. 
This provides a differentiable proxy to the a-DCF that can be optimized through standard gradient-based approaches. We adopt similar strategy in the present work, with further detail provided in Section \ref{sec:proposed-approach}.

% Earlier Version of Sec. III
%\input{existing_studies_for_SASV}
% Cemal's Version
\section{Existing SASV Approaches}\label{sec:existing-sasv-models}

Recent research on SASV has explored various strategies to combine ASV and CM tasks. %We provide a compact summary of existing approaches in Table~\ref{tab:existing-systems-summary}. 
The approaches can be broadly categorized into two main directions:

\textbf{1) End-to-end approaches:} End-to-end systems map a raw speech waveform directly to a single SASV detection score. The core idea is to train a unified network to simultaneously learn speaker-discriminative features and spoof-related artifacts \cite{kang22_interspeech}. The primary advantage of this approach is its potential for high performance by jointly optimizing all system components toward a single objective, such as the a-DCF~\cite{shim2024adcf}. In these systems, the entire pipeline—from feature extraction to final scoring—is tightly coupled and optimized as a single unit.

\begin{figure*}[t]
  \centering
  \includegraphics[width=0.9\textwidth]{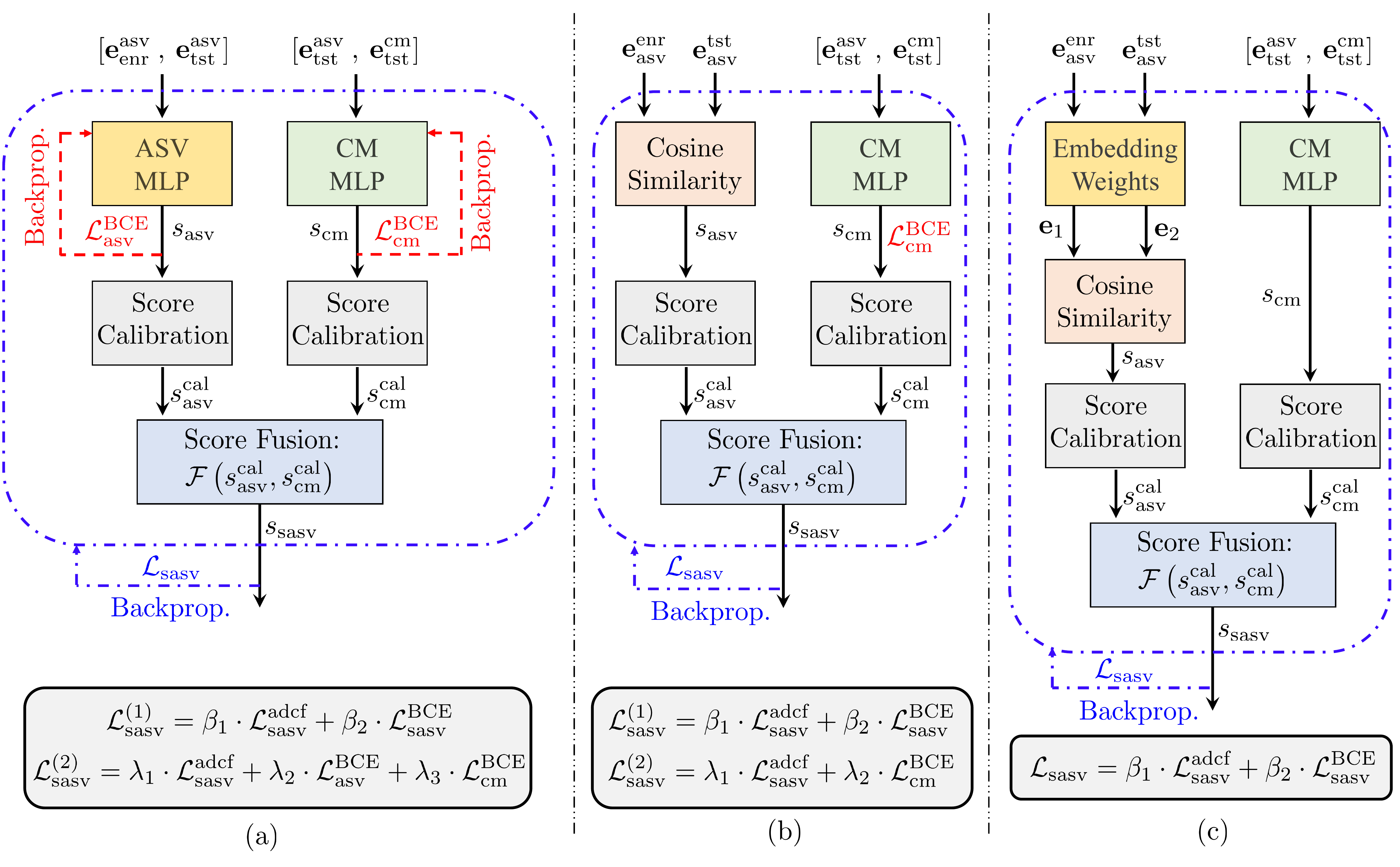}
  \caption{Illustration of the three proposed modular SASV architectures. Each system comprises four components: (i) an ASV branch for extracting speaker embeddings and computing the ASV score ($s_{\mathrm{asv}}$), (ii) a CM branch for detecting spoofed speech via the CM score ($s_{\mathrm{cm}}$), (iii) a score fusion module that integrates the ASV and CM scores, and (iv) an optimization strategy that either jointly or separately tunes the system components.}
  \label{fig:your_label}
\end{figure*}

However, this tight coupling %comes at the cost of 
can be at odds with interpretability and explaining decisions. If an end-to-end system produces a false acceptance (e.g., accepting a spoofed voice as genuine), it is difficult to determine whether the error was caused by a failure in detecting spoofing artifacts or by misclassifying the speaker. The final score results from complex non-linear interactions within a single network, making post-hoc error attribution effectively impossible. 
This lack of transparency can be a critical limitation for high-security applications where understanding the source of failures is essential.

\textbf{2) Combining outputs of ASV and CM subsystems:} The second, and more common SASV strategy, trains ASV and CM subsystems separately and combines their outputs at different levels: (i) embeddings, (ii) decision scores, or (iii) hard decisions~\cite{Shim2022}. Maintaining separate subsystems preserves modularity, allowing them to be replaced or upgraded independently.

Score-level fusion is widely used, with either simple linear strategies (e.g., averaging or weighted summation) or non-linear methods (e.g., logistic regression, Gaussian back-end~\cite{todisco18_interspeech} or MLP-based fusion) to produce the final SASV decision. Embedding-level fusion is a more sophisticated approach: pre-trained ASV and CM models generate high-dimensional vectors---'speaker' and 'spoof' embeddings---which are typically concatenated and processed by a separate back-end classifier. This strategy, explored in the SASV2022 challenge baselines, leverages rich information from both subsystems while allowing for multi-level fusion. Hard decision fusion, in contrast, applies a logical AND rule, rejecting a trial if either subsystem %flags it as non-genuine.
rejects it. This last approach has been the strategy adopted in ASVspoof challenges~\cite{tdcf_kinnunen2020}.

Despite providing improved %more 
modularity compared to fully end-to-end systems, embedding-level fusion still limits diagnosability. Once ASV and CM embeddings are combined, the contributions of each subsystem are entangled and the classifier cannot provide a clear breakdown of which component influenced the decision. As a result, failures cannot be directly attributed to ASV or CM subsystems, which complicates system analysis, debugging, and accountability for decision errors.

Overall, existing SASV systems primarily differ in (i) how ASV and CM subsystems are integrated, (ii) whether fusion occurs at the embedding or score level and (iii) whether subsystems are trained independently or jointly. These trends motivate the development of modular, non-linear, and jointly optimized SASV frameworks, which form the basis of our proposed approach. Table~\ref{tab:sasv_summary} summarizes representative methods, highlighting their ASV/CM backbones and fusion strategies.

\begin{table*}[ht]
\centering
\scriptsize
\caption{Summary of SASV systems. 
Each entry lists the \textit{ASV System}, which specifies the backbone model used for the ASV task; the \textit{CM System}, which indicates the countermeasure architecture employed for the CM task; and the \textit{SASV architecture}, which describes how these ASV and CM subsystems are combined.}
\label{tab:sasv_summary}

\resizebox{\linewidth}{!}{%
\begin{tabular}{|l|l|l|l|}
\hline
\textbf{Paper} & \textbf{ASV System} & \textbf{CM System} & \textbf{SASV Architecture} \\
\hline

\multicolumn{4}{|c|}{\textbf{End-to-End Structure}} \\
\hline
\cite{kang22_interspeech} & Unified SASV network & Same as ASV system & \multirow{4}{*}{End-to-end framework} \\
\cite{teng22_interspeech} & A spoof-aggregated SASV network & Same as ASV system & \\
\cite{Zhang2022sasv} & Unified ECAPA-TDNN model & Same as ASV system & \\
\cite{Alenin2022} & Unified SASV network with CM subnetwork & Same as ASV system & \\
\hline

\multicolumn{4}{|c|}{\textbf{Modular Structure}} \\
\hline
\cite{wang24l_interspeech} & ECAPA-TDNN & AASIST & \multirow{28}{*}{ASV--CM outputs fusion} \\
\cite{Zhang2022} & ECAPA-TDNN & AASIST &  \\
\cite{Wang2022b} & ResNet based models, ECAPA-TDNN & AASIST based models & \\
\cite{zhang22w_interspeech} & ECAPA-TDNN & AASIST & \\
\cite{rohdin24_asvspoof} & ResNet variants & ResNet18 or SSL models & \\
\cite{martindonas24_asvspoof} & TitaNet & Wav2Vec2, WavLM & \\
\cite{villalba24_asvspoof} & Modified ResNet100 & Modified ResNet34 & \\
\cite{aliyev24_asvspoof} & ResNet variants & WavLM based ensembles & \\
\cite{lin22_interspeech} & ECAPA-TDNN & AASIST variant & \\
\cite{wang22ea_interspeech} & ResNet based ensembles, ECAPA & AASIST variant & \\
\cite{tran24_asvspoof} & ResNet variants & ResNet34 & \\
\cite{okhotnikov24_asvspoof} & ResNet100 & SSL Transformers + CNN models & \\
\cite{duroselle24_asvspoof} & ResNet34 & ResNet+AASIST+autoencoder ensembles & \\
\cite{zhang22f_interspeech} & ResNet variants, ECAPA & AASIST variants & \\
\cite{mun23_interspeech} & Unified SASV extractor & Same as ASV system & \\
\cite{choi22b_interspeech} & ResNet34, Res2Net & AASIST & \\
\cite{Lee2022} & ECAPA-TDNN & Wav2Vec & \\
\cite{heo22_interspeech} & ECAPA-TDNN & AASIST & \\
\cite{Wu2022e} & ResNet34, ECAPA-TDNN, MFA-Conformer & AASIST variants & \\
\cite{chen24_asvspoof} & ResNet242 & AASIST, RawNet2, W2V2, Res2Net & \\
\cite{asali25_interspeech} & ECAPA-TDNN & AASIST & \\
\cite{buker25_interspeech} & ECAPA-TDNN & AASIST & \\
\cite{li25h_interspeech} & ResNet34 & ResNet variant & \\
\cite{kurnaz24_asvspoof} & ECAPA-TDNN, WavLM & AASIST & \\
\cite{11111888} & ECAPA-TDNN & AASIST variant & \\
\hline
Ours & ECAPA-TDNN, WavLM-TDNN, ReDimNet & AASIST variants & Unified SASV system with non-linear fusion \\
\hline
\end{tabular}}
\end{table*}

\section{Proposed Approach}\label{sec:proposed-approach}

As illustrated in Fig.~\ref{fig:your_label}, we propose three modular SASV systems that jointly address the ASV and CM tasks within a unified optimization framework, by %The architectures are designed to exploit ASV embeddings and anti-spoofing cues (CM embeddings) in a complementary manner, thereby enhancing system robustness against spoofing attacks while preserving high speaker discriminability. 
leveraging embeddings specialized in speaker and spoof detection.
%The proposed modular architectures are illustrated in Fig.~\ref{fig:your_label}. 
Each architecture consists of three key elements: %for obtaining the final SASV score: 
(i) \emph{an ASV branch}, which computes the ASV score ($s_{\mathrm{asv}}$) for a pair of %enrollment and test ASV 
speaker embeddings ($\mathbf{e}_\mathrm{asv}^{\mathrm{enr}}$ and $\mathbf{e}_\mathrm{asv}^{\mathrm{tst}}$), (ii) \emph{a CM branch}, which %performs the spoofing detection task by 
computes the realness score ($s_{\mathrm{cm}}$) for a given CM embedding ($\mathbf{e}_{\mathrm{cm}}^\mathrm{tst}$) of a test utterance, (iii) a \emph{fusion module}, which integrates the ASV and CM scores. In addition to these architectural considerations, (iv) \emph{optimization strategy} is another critical choice.
%which either jointly or separately optimizes the individual components of the overall SASV system. 
In the following, % subsections, 
we provide further detail on all these elements.

%a brief description of each component in the proposed framework.

\subsection{ASV Branch: Embedding Weighting and Similarity}
The ASV branch %(located on the left side 
(left-most side of each system in Fig.~\ref{fig:your_label}) performs %the speaker verification task 
speaker comparison by computing the ASV score $s_{\mathrm{asv}}$ from a pair of deep speaker embeddings extracted from the enrollment ($\mathbf{X}_\mathrm{enr}$) and test ($\mathbf{X}_\mathrm{tst}$) utterances,
\begin{equation}
    \begin{aligned}
    \mathbf{e}_{\text{enr}}^{\text{asv}} & = \mathrm{emb}_\mathrm{ASV} (\mathbf{X}_\mathrm{enr}) \in \mathbb{R}^{1 \times D_\text{asv}}\\\mathbf{e}_{\text{tst}}^{\text{asv}} &= \mathrm{emb}_\mathrm{ASV} (\mathbf{X}_\mathrm{tst}) \in \mathbb{R}^{1 \times D_\text{asv}},    
    \end{aligned}
    %\mathbf{e}_{\text{enr}}^{\text{asv}} = \mathrm{emb}_\mathrm{ASV} (\mathbf{X}_\mathrm{enr}) \in \mathbb{R}^{1 \times D_\text{asv}}, \: \mathbf{e}_{\text{tst}}^{\text{asv}} = \mathrm{emb}_\mathrm{ASV} (\mathbf{X}_\mathrm{tst}) \in \mathbb{R}^{1 \times D_\text{asv}},
\end{equation}
%$\mathbf{e}_{\text{enr}}^{\text{asv}} = \mathrm{emb}_\mathrm{ASV} (\mathbf{X}_\mathrm{enr})$ and $\mathbf{e}_{\text{tst}}^{\text{asv}} \mathrm{emb}_\mathrm{ASV} (\mathbf{X}_\mathrm{tst})$ 
%denote the deep speaker embeddings extracted from enrollment and test utterances, respectively, 
where $\mathrm{emb}_\mathrm{ASV}(\cdot)$ denotes a pre-trained ASV embedding extractor and $D_\text{asv}$ is the dimensionality of the speaker embeddings. The speaker similarity score $s_{\mathrm{asv}}$ is computed %in the proposed framework 
using one of three alternative strategies:

\begin{itemize}
    \item \textbf{MLP-based similarity (Fig.~\ref{fig:your_label} (a)):} The enrollment and test embeddings are concatenated, $\mathbf{e}^\mathrm{asv}=\left[ \mathbf{e}_\mathrm{enr}^\mathrm{asv} \: , \: \mathbf{e}_\mathrm{tst}^\mathrm{asv} \right]$. An MLP-based %ASV binary classifier 
    speaker comparator $f$
    %$f_{\theta_\mathrm{asv}}$, parameterized by $\theta_\mathrm{asv}$, 
    then outputs score %the ASV score, 
    $s_\mathrm{ASV} = f_{\theta_\mathrm{asv}}(\mathbf{e}^\mathrm{asv})$, where 
    %$f_{\theta_\mathrm{asv}}$ 
    $\theta_\mathrm{asv}$ denotes the %MLP 
    parameters.
    %This ASV classifier 
    This MLP aims to discriminate target and non-target trials. %using embedding pairs from the training set using the binary cross-entropy (BCE) loss 
    It is trained using binary cross-entropy (BCE). %as indicated by the red dashed line in Fig.~\ref{fig:your_label} (a).
    
    \item \textbf{Cosine similarity (Fig.~\ref{fig:your_label} (b)):} The cosine similarity between $\mathbf{e}_\mathrm{enr}^\mathrm{asv}$ and $\mathbf{e}_\mathrm{tst}^\mathrm{asv}$ is directly used as the ASV score,  
    \[
    s_{\mathrm{asv}} = \frac{\mathbf{e}_{\text{enr}}^{\text{asv}} \cdot {\mathbf{e}_{\text{tst}}^{\text{asv}^\mathrm{T}}}}{\|\mathbf{e}_{\text{enr}}^{\text{asv}}\| \|\mathbf{e}_{\text{tst}}^{\text{asv}}\|}.  
    \]  
    \item %\textbf{Embedding weighting with cosine similarity (Fig.~\ref{fig:your_label} (c)):} 
    \textbf{Weighted cosine similarity (Fig.~\ref{fig:your_label} (c)):} 
    A learnable version of cosine score, where
    the enrollment and test embeddings are first %element-wise 
    weighted element-wise: %using a learnable weighting vector to obtain refined embeddings:
    \[
    \mathbf{e}_1 = \mathbf{w}_{\text{asv}} \odot \mathbf{e}_{\text{enr}}^{\text{asv}}, \quad
    \mathbf{e}_2 = \mathbf{w}_{\text{asv}} \odot \mathbf{e}_{\text{tst}}^{\text{asv}}
    \]
    where $\mathbf{w}_{\text{asv}} \in \mathbb{R}^{1 \times D_{\mathrm{asv}}}$ denotes a shared learnable parameter vector and $\odot$ denotes element-wise (Hadamard) product. %multiplication. 
    This weighting operation allows the model to learn the relative importance of embedding dimensions for speaker discrimination. The ASV score is then %obtained by 
    computed as the cosine similarity of %similarity 
    $\mathbf{e}_1$ and $\mathbf{e}_2$.
    %the weighted embeddings:
    %\[
    %s_{\text{asv}} = \frac{\mathbf{e}_1 \cdot \mathbf{e}_2}{\|\mathbf{e}_1\| \|\mathbf{e}_2\|}
    %\]
\end{itemize}
To improve score interpretability and to enable effective fusion with the CM score, the resulting raw ASV score $s_{\text{asv}} \in \mathbb{R}$ is passed through a learnable %linear 
affine calibration layer to obtain %the 
calibrated ASV score: %so that it can be effectively treated as LLR as defined in~\eqref{eq:asv_llr}:
\begin{equation}
\ell^{\text{tar.bon}}_{\text{non.bon}}(\vec{X}) \approx s_{\text{asv}}^{\text{cal}} :=  w_0^{\text{asv}} + w_1^{\text{asv}} s_{\text{asv}},
\end{equation}
where $w_0^{\text{asv}}, w_1^{\text{asv}} \in \mathbb{R}$ denote scalar calibration parameters. The final score can be effectively treated as an LLR, as in~\eqref{eq:asv_llr}. The use of an affine transform for ASV score calibration is particularly effective here. Since cosine similarity scores already reside within a bounded range, we assume that they are often somewhat linearly related to log-likelihood ratios even before calibration.

\subsection{CM Branch: Score from Joint Embeddings}
To integrate spoofing awareness, the system employs a CM classifier %located on the 
(right-most side of each system in Fig.~\ref{fig:your_label}) that operates on the concatenated ASV and CM embeddings derived from the test utterance $\vec{X}_\text{tst}$. Concretely, let $\mathbf{e}_{\text{tst}}^{\text{cm}} = \text{emb}_\text{CM} (\vec{X}_\text{tst})\in \mathbb{R}^{1 \times D_\text{cm}}$ denote the CM-specific embedding of the test utterance, where $\text{emb}_\text{CM}(\cdot)$ is a pre-trained CM embedding extractor network and $D_\text{cm}$ is the dimensionality of the CM embedding. $\mathbf{e}_{\text{tst}}^{\text{cm}}$ is then concatenated with the ASV embedding $\mathbf{e}_{\text{tst}}^{\text{asv}} \in \mathbb{R}^{1 \times D_\text{asv}}$ to form a joint representation:
\begin{equation}
\mathbf{e}_{\text{fused}} = [\mathbf{e}_{\text{tst}}^{\text{asv}} \, ; \, \mathbf{e}_{\text{tst}}^{\text{cm}}] \in \mathbb{R}^{1 \times (D_\text{asv} + D_\text{cm})}
\end{equation}
This joint representative vector is passed to a CM classifier, implemented as an %multi-layer perceptron (MLP) 
MLP parameterized by $\mathbf{\theta}_\text{cm}$ ($f_{\mathbf{\theta}_\text{cm}}$), which produces a scalar CM score:
\begin{equation}
s_{\text{cm}} = f_{\mathbf{\theta}_\text{cm}}(\mathbf{e}_{\text{fused}})
\end{equation}
Similar to the ASV branch, a separate calibration step is applied to the CM score to approximate the LLR defined in~\eqref{eq:cm_llr}:
\begin{equation}
\ell^{\text{tar.bon}}_{\text{spf}}(\vec{X}) \approx s_{\text{cm}}^{\text{cal}} := w_0^{\text{cm}} + w_1^{\text{cm}} s_{\text{cm}}
\end{equation}
where $w_0^{\text{cm}}, w_1^{\text{cm}} \in \mathbb{R}$ are scalar learnable calibration parameters used to calibrate the raw CM score. Applying score calibration to both ASV and CM scores ensures that both scores are on a compatible scale prior to score fusion and they both can be treated as LLRs. The raw logits produced by the MLP-based CM classifier lack a probabilistic scale and are often poorly calibrated. The affine calibration layer therefore transforms these arbitrary logits into calibrated LLRs, placing them on a compatible scale with the ASV scores prior to the non-linear fusion defined in Section~\ref{sec:scorefusionandjointdecision}.

\subsection{Score Fusion and Joint Decision}
\label{sec:scorefusionandjointdecision}
Once the calibrated ASV and CM scores ($\ell^{\text{tar.bon}}_{\text{non.bon}}(\vec{X})$ and $\ell^{\text{tar.bon}}_{\text{spf}}(\vec{X})$) are computed, %from the ASV and CM branches of the proposed SASV model, 
the final SASV score $s_\text{sasv}\in \mathbb{R}$ is obtained by fusing them: %two scores using either a linear combination or a more expressive non-linear fusion model: 
\begin{equation}
s_{\text{sasv}} = \mathcal{F}\left( \ell^{\text{tar.bon}}_{\text{non.bon}}(\vec{X}), \ell^{\text{tar.bon}}_{\text{spf}}(\vec{X}) \right)
\end{equation}
where $\mathcal{F}$ denotes a generic fusion function.  %Both simple 
In our experiments, we consider both linear%fusion as in
~\eqref{eq:linear-fusion-of-LLRs} and %a more expressive 
non-linear~\eqref{eq:basic-nonlinear-fusion} approaches. %are explored as the fusion function. 
The fused score represents a joint decision that accounts for both speaker identity and spoofing status.

\subsection{Joint Optimization and Loss Function}

The final SASV decision score $s_{\text{sasv}}$ represents the system's confidence that the test utterance is both from the claimed speaker and bonafide (i.e., not spoofed). To supervise this joint objective, a loss function is adopted and the entire system, including embedding projections, the CM classifier, calibration layers, and fusion function, is trained end-to-end using this loss function. Gradients from the loss are backpropagated throughout the network, encouraging both branches to optimize jointly toward spoofing-robust verification performance. To adopt the loss function, each training sample consists of an enrollment-test pair is labeled as:
\begin{itemize}
\item Positive ($y_\text{sasv}=1$): Same speaker and bonafide test utterance
\item Negative ($y_\text{sasv}=0$): Either spoofed or from a different speaker.
\end{itemize}
In our experiments, %binary cross-entropy (BCE) 
we consider BCE %loss is utilized as the 
as our baseline loss: %function which is defined as:
\begin{equation}
\mathcal{L}_{\text{sasv}}^{\text{BCE}} = - \left[ y_{\text{sasv}} \cdot \log s_{\text{sasv}} + (1 - y_{\text{sasv}}) \cdot \log(1 -s_{\text{sasv}}) \right].
\end{equation}
%where $y_{\text{sasv}} \in \{0, 1\}$ is the ground truth SASV label. The BCE loss 
BCE encourages the model to produce higher SASV scores for bonafide target trials and lower scores for either spoofing or non-target speaker trials. Nonetheless, 
%Although BCE loss is widely used cost function to optimize the model parameters for the vast majority of the binary classification problems, 
BCE is arguable a suboptimal choice for the SASV task %. This is because SASV performance is assessed through the tree error rates, 
that must trade between the possibly conflicting error rate terms
$P_{\text{miss}}^{\text{tar}}, P_{\text{fa}}^{\text{non}}$ and $P_{\text{fa}}^{\text{spf}}$. %The BCE loss does not explicitly optimize. 
Therefore, we incorporate the a-DCF loss%(defined in~\cite{kurnaz2024optimizing})
, $\mathcal{L}_{\text{sasv}}^{\text{adcf}}$, which is %more 
aligned closer with SASV task. %evaluation criterion and offers improved robustness against spoofing attacks. 
This loss estimates (differentiable versions of) the three error rates from $s_{\text{sasv}}$ at a threshold $\tau_{\text{sasv}}$ and combines them into an a-DCF objective using a-DCF loss as described in Sec.~\ref{sec:adcf_loss} and originally proposed in~\cite{kurnaz2024optimizing}. 

In addition to BCE and a-DCF losses, to facilitate stable optimization and to improve convergence, auxiliary BCE losses are introduced for the ASV and CM branches: 
\begin{align}
\mathcal{L}_{\text{asv}}^{\text{BCE}} &= - \left[ y_{\text{asv}} \cdot \log s_{\text{asv}}^{\text{cal}} + (1 - y_{\text{asv}}) \cdot \log(1 - s_{\text{asv}}^{\text{cal}}) \right] \\
\mathcal{L}_{\text{cm}}^{\text{BCE}} &= - \left[ y_{\text{cm}} \cdot \log s_{\text{cm}}^{\text{cal}} + (1 - y_{\text{cm}}) \cdot \log(1 - s_{\text{cm}}^{\text{cal}}) \right]
\end{align}
where $y_{\text{asv}} \in \{0, 1\}$  indicates whether the enrollment and test embeddings are from the same speaker and $y_{\text{cm}} \in \{0, 1\}$   indicates whether the trial is bonafide or spoofed. These auxiliary objectives allow the individual branches to learn task-specific discriminative features prior to score fusion. The final loss is a combination of the main SASV loss and auxiliary losses, enabling joint optimization while preserving the modular design. We consider two variants of the final training loss:
\begin{equation}
\mathcal{L}_\text{sasv}^{(1)} = \beta_1 \cdot \mathcal{L}_{\text{sasv}}^{\text{adcf}} + \beta_2 \cdot \mathcal{L}_{\text{sasv}}^{\text{BCE}} 
\label{eq:sasv_loss1}
\end{equation}
\begin{equation}
\mathcal{L}_\text{sasv}^{(2)} = \lambda_1 \cdot \mathcal{L}_{\text{sasv}}^{\text{adcf}} + \lambda_2 \cdot \mathcal{L}_{\text{asv}}^{\text{BCE}} + \lambda_3 \cdot \mathcal{L}_{\text{cm}}^{\text{BCE}}
\label{sasv_loss2}
\end{equation}
where all weighting coefficients are predefined fixed hyperparameters selected prior to training and kept constant throughout optimization; in our implementation, they are set to unity for equal contribution of each component. The final loss (either $\mathcal{L}_\text{sasv}^{(1)}$ or $\mathcal{L}_\text{sasv}^{(2)}$) is backpropagated jointly through both the ASV and CM branches, allowing the system to learn unified representations and decision boundaries while maintaining a modular structure that supports component-level evaluation and tuning. This joint optimization approach provides a balance between integration and interpretability, making it a practical and robust solution for spoof-aware speaker verification.

\section{Experimental Setup}

\subsection{Datasets}

We conducted our experiments using the ASVspoof 5 dataset~\cite{Wang2024_ASVspoof5}, the latest release in the ASVspoof challenge series. Compared to earlier versions, this dataset is larger and incorporates more advanced %state-of-the-art techniques, 
attack algorithms, including various TTS, VC, and adversarial attacks. It also offers greater speaker diversity and a wider range of acoustic conditions. The dataset consists of %two tracks: 
evaluation protocols for both %(1) Track 1 for 
deepfake detection (Track 1), and %(2) Track 2 for 
SASV (Track 2). %Since our study focuses 
We focus exclusively on the latter. %SASV task (Track 2). %we used only Track 2. 

ASVspoof 5 organizes the dataset into three subsets: training, development, and evaluation. The training set includes bonafide utterances and spoofed samples generated using eight distinct TTS methods. The development set contains bonafide utterances along with spoofed samples created using five TTS and three VC systems. The evaluation set introduces a broader range of sixteen spoofing attacks, comprising TTS, VC, and adversarial attacks (AT). Moreover, both the bonafide and spoofed samples %the %evaluation set applies 
%data is 
are processed through various compressor-decompressor (codec) techniques %to both bonafide and spoofed data, 
employing varied bitrates and sampling frequencies to mimic real-world transmission and audio storage conditions. A summary of the speaker counts, utterance distributions, and %spoofing attack types across all three subsets 
attacks is provided
in Table~\ref{tab:asvspoof5_dataset}.

\begin{table}[ht]
\centering
\caption{ASVspoof 5 data statistics. TTS: text to speech, VC: voice conversion, AT: adversarial attack.}
\label{tab:asvspoof5_dataset}
\begin{adjustbox}{width=\linewidth}
\begin{tabular}{|c|c|cc|cc|}
\hline
\textbf{Subset} & \textbf{Att. Type} & \multicolumn{2}{c|}{\textbf{\# Utterances}} & \multicolumn{2}{c|}{\textbf{\# Speakers}} \\
                &                    & \textbf{Bonafide} & \textbf{Spoof} & \textbf{Female} & \textbf{Male} \\ \hline
Trn.  & TTS (8)           & 18,797  & 163,560  & 196  & 204  \\
Dev.  & TTS (5) / VC (3)  & 31,334  & 109,616  & 392  & 393  \\
Eval. & TTS (6) / VC (3) / AT (7) & 138,688 & 542,086 & 370  & 367  \\ \hline
\end{tabular}
\end{adjustbox}
\end{table}

%\subsection{Experimental Setup}
\subsection{SASV Approaches}

%To establish a fair comparison, we implement a set of baseline and proposed variants. We first systematically analyze existing approaches using score fusion\cite{wang24l_interspeech} and modular fusion\cite{kurnaz24_asvspoof}. 
%We summarize the variants below:
We consider the following four SASV approaches.
\begin{itemize}
\item \textbf{Score fusion}~\cite{wang24l_interspeech}: The SASV system takes ASV and CM scores and calibrates them with LLR ~\cite{wang24l_interspeech}. It then combines the calibrated scores either linearly (using \eqref{eq:linear-fusion-of-LLRs}) or nonlinearly (using \eqref{eq:basic-nonlinear-fusion}) to produce the final SASV score $s_{\text{sasv}}$.

%Uses ASV embeddings from a publicly available ASV system with cosine similarity and CM system scores from a publicly available baseline. The two scores with LLR calibration~\cite{wang24l_interspeech} are combined through linear and non-linear fusion to obtain the SASV score $s_{\text{sasv}}$.

\item \textbf{MLP-based classification with score fusion}: CM and ASV embeddings pass through separate MLPs (Fig.~\ref{fig:your_label}(a), excluding the score calibration and joint optimization part). Each MLP is optimized with BCE for its respective task—spoof detection or speaker verification. We then fuse the calibrated LLR scores~\cite{wang24l_interspeech} from both MLPs to obtain the final SASV score $s_{\text{sasv}}$.

\item \textbf{Joint optimization of ASV and CM MLP}~\cite{kurnaz24_asvspoof}: Both ASV and CM branches employ MLP-based classifiers, as illustrated in Fig.~\ref{fig:your_label}(a). They are jointly optimized with a shared trainable calibration layer, following the formulations in Equations~\eqref{eq:asv_llr} and~\eqref{eq:cm_llr}.  

\item \textbf{Cosine similarity for ASV with MLP-based CM}: Shown in Fig.~\ref{fig:your_label}(b), this variant reflects the different nature of the tasks; ASV as detection and CM as classification. Cosine similarity generates ASV scores between enrollment and test embeddings, which are passed through a trainable calibration layer. In parallel, the CM branch uses a trainable MLP classifier with its own calibration layer. Here, only the CM branch is trainable, while the ASV branch remains fixed except for calibration.  

\end{itemize}

Finally, guided by the insights from this analysis, we propose a \textbf{unified SASV system} that jointly addresses speaker and spoof detection. As illustrated in Fig.~\ref{fig:your_label}(c), the ASV branch employs a domain-inspired \emph{embedding weighting} scheme with \emph{cosine similarity}. In contrast, the CM branch uses CM embeddings through an \emph{MLP-based classifier}. The system is optimized end-to-end using non-linear score fusion with calibration and jointly trained with both a-DCF and BCE objectives.

\subsection{ASV and CM Embedding Extractor}
We used three publicly available ASV embedding extractors in our study: (1) Emphasized Channel Attention, Propagation and Aggregation Time Delay Neural Network (ECAPA-TDNN)\footnote{https://github.com/TaoRuijie/ECAPA-TDNN/ (accessed \myshortdate\today)}\cite{Desplanques_2020}, (2) WavLM-TDNN\footnote{https://huggingface.co/microsoft/wavlm-base-sv (accessed \myshortdate\today)}\cite{9814838} and (3) ReDimNet\footnote{https://github.com/IDRnD/redimnet (accessed \myshortdate\today)}\cite{yakovlev24_interspeech}. We selected ECAPA-TDNN for its widespread adoption in the speaker verification literature, WavLM-TDNN for its use of self-supervised representations with stronger cross-domain generalization, and ReDimNet for its competitive performance with lightweight architecture. Similarly, we employed two CM embedding extractors: (1) Audio Anti-Spoofing using Integrated Spectro-Temporal graph (AASIST)\cite{jung2022aasist} and (2) its self-supervised variant SSL-AASIST\cite{tak22_odyssey}. AASIST, a graph neural network–based classifier, has been widely adopted in the spoofing detection literature, while SSL-AASIST leverages self-supervised learning to enhance generalization. SSL-AASIST employs a self-supervised wav2vec 2.0 XLS-R\footnote{https://github.com/facebookresearch/fairseq/tree/main/examples/wav2vec/xlsr (accessed \myshortdate\today)} front-end, which has been pre-trained on 436k hours of unlabeled audio drawn from diverse corpora, including VoxPopuli, MLS, CommonVoice, BABEL, and VoxLingua107, and therefore must be considered under the \emph{Open Condition} rules of the ASVspoof5 challenge. This front-end is then fine-tuned jointly with the AASIST back-end classifier using the ASVspoof5 training set. During fine-tuning, a fully connected layer with $128$ output dimensions is added after the wav2vec 2.0 features, and audio segments of approximately $4$ seconds are used as input with various data augmentation strategies (noise, reverberation, etc.). For optimization, SSL-AASIST uses a smaller batch size ($14$) and a reduced learning rate of $1 \times 10^{-6}$ to avoid over-fitting. The ASV and CM models contain approximately $14.7$M (ECAPA-TDNN), $94.7$M (WavLM-TDNN), $15$M (ReDimNet), $297$K (AASIST), and $15$M (SSL-AASIST) parameters. While larger models tend to yield stronger performance, they also incur higher computational cost and slower inference speed. 

\subsection{MLP-based Classifier and Training details}

In our earlier work~\cite{kurnaz24_asvspoof}, we adopted the ``Baseline-2'' architecture~\cite{Shim2022} from the SASV challenge as a placeholder for embedding fusion. This architecture consists of three hidden layers with $256$, $128$, and $64$ nodes, respectively. However, the choice of these hyperparameters was largely heuristic or based on arbitrary settings, without consideration of optimality for the joint training of ASV and CM. To address this limitation, we performed a more systematic optimization of the architecture, learning rate, and batch size using a Bayesian search approach, as described in~\cite{akiba2019optuna}. This search iteratively evaluated candidate configurations by sampling learning rates in the range $[10^{-5}, 10^{-2}]$ on a log scale, batch sizes from $64$ to $1024$ in steps of $64$, and architectures spanning $2$--$6$ hidden layers with per-layer widths from $64$ to $512$ (step $32$). Table~\ref{tab:score_fusion_arch} summarizes the architecture, learning rate, and batch size for both the baseline-2 setup and the Bayesian-optimized configuration.

% In the initial setup, each classifier follows the architecture proposed as Baseline~2 in the SASV 2022 challenge, consisting of three hidden layers with $256$, $128$, and $64$ nodes, respectively. 
% To achieve the lowest SASV-EER, both ASV and CM classifiers were optimized via Bayesian search, which identified a two-hidden-layer configuration with $384$ and $160$ nodes as optimal (Table~\ref{tab:score_fusion_arch}). 
% This symmetrical architecture ensures that the ASV and CM networks have identical layer and node counts, maintaining architectural consistency while processing task-specific embeddings.

\begin{table}[H]
\centering
\caption{Classifier architectures for score-level SASV. 
\textit{Baseline-2 (SASV2022)} denotes the reference model provided by the SASV2022 challenge~\cite{Jung2022} organizers, 
while \textit{Proposed work} presents the optimized architecture obtained through Bayesian optimization in this study.}
\begin{tabular}{|l|c|c|}
\hline
\textbf{Parameter} & \textbf{Baseline-2 (SASV2022)} & \textbf{Proposed work} \\
\hline
Hidden Layers       & 3 Layers        & 2 Layers        \\
Node Sizes          & 256, 128, 64    & 384, 160        \\
Batch Size          & 1024            & 192             \\
Learning Rate       & 0.0001          & 0.000861        \\
\hline
\end{tabular}
\label{tab:score_fusion_arch}
\end{table}

%For the ASV branch, two different embedding extractors are employed: the \textbf{ECAPA-TDNN} model, a widely adopted deep speaker verification framework, and the \textbf{WavLM-TDNN} model, which integrates transformer-based contextual representations with a TDNN backend for enhanced verification accuracy. 
%For the CM branch, embeddings are obtained from the \textbf{AASIST} model, known for its robustness against spoofed speech, and its self-supervised learning variant \textbf{SSL-AASIST}, which incorporates SSL-based features to improve detection of unseen spoofing attacks.

%\subsection{Training Details}
Except for conventional score fusion (which does not require optimization), we trained all systems for $100$ epochs. For the unified SASV system (Fig.~\ref{fig:your_label}(c)), the embedding weight parameters $w_{\mathrm{asv}}$ are initialized randomly and trained from scratch. We chose the optimal model checkpoint for evaluation by monitoring the min a-DCF on the development set, where the cost values $C_{\mathrm{miss}}^{\mathrm{tar.bon}}$, $C_{\mathrm{fa}}^{\mathrm{non.bon}}$, and $C_{\mathrm{fa}}^{\mathrm{spf}}$ in the a-DCF calculation are set as $1$, $10$, and $20$, respectively, while the prior values $\pi_{\mathrm{tar}}$, $\pi_{\mathrm{non}}$, and $\pi_{\mathrm{spf}}$ are set as $0.9$, $0.05$, and $0.05$. We optimized the models using either adaptive moment estimation (Adam)~\cite{kingma2015adam} or the vanilla stochastic gradient descent (SGD) optimizer as implemented in PyTorch~\cite{sutskever2013importance}.

% \textcolor{blue}{
% Beyond the baseline configurations, we also evaluated the integration 
% of margin-based objectives to enhance speaker separability. 
% For the experiments involving margin-based objectives, the cosine embedding loss 
% was incorporated to both the MLP--MLP (Fig.~\ref{fig:your_label}(a)) and weighted cosine + MLP 
% (Fig.~\ref{fig:your_label}(c)) architectures. Consistent with our findings on joint optimization, 
% these models were initialized randomly and trained using the SGD optimizer. 
% The training was supervised by a combined objective consisting of the 
% soft a-DCF loss ($\mathcal{L}_{\text{sasv}}^{\text{adcf}}$), 
% the SASV BCE loss ($\mathcal{L}_{\text{sasv}}^{\text{BCE}}$), 
% and the cosine embedding loss ($\mathcal{L}_{\text{asv}}^{\text{cos}}$).}

\subsection{Evaluation Metrics}

%In this study, three main metrics are employed: 
We employ three performance metrics: %the architecture-agnostic detection cost function 
a-DCF~\cite{shim2024adcf}, the speaker verification equal error rate (SV-EER), and the spoof equal error rate (SPF-EER). The parameters of a-DCF~\eqref{adcf_old} 
%is used to evaluate spoofing-aware ASV systems by combining the costs and priors of three types of errors: missed detections of target speakers, false acceptances of non-target speakers, and false acceptances of spoofed trials. 
are the same as used in the optimization (see above). We consider both the `minimum' a-DCF (obtained by selecting $\tau_{\mathrm{sasv}}$ on evaluation data that minimizes~\eqref{adcf_old}) and the `actual' a-DCF (obtained by selecting a single uniform threshold ($\tau^{\mathrm{dev}}_{\mathrm{sasv}}$) on pooled development set).
%In practice, a-DCF is computed by sweeping a decision threshold over system scores and selecting the operating point that minimizes this cost. 
The SV-EER corresponds to the threshold where the miss rate of target speakers equals false alarm rate of non-target speakers. SPF-EER is obtained %analogously 
similarly by equating the miss rate of target speakers with the false alarm rate of spoofing attacks.

\definecolor{hifiganbg}{RGB}{220,240,255}   % light blue
\definecolor{wavconcatbg}{RGB}{255,235,220} % light orange
\definecolor{bigvganbg}{RGB}{235,220,255}   % light purple
\definecolor{advattbg}{RGB}{220,255,220}    % light green  

\section{Experimental Results}
\label{sec:experimental_results}

\subsection{Score Fusion vs.\ MLP Classifier with Score Fusion}

Table~\ref{tab:score_vs_mlp} compares conventional score fusion with MLP-based score fusion under both linear and non-linear fusion strategies on the development set. Two trainable MLPs are optimized separately for ASV and CM using their respective objectives, and the resulting scores are fused in the same way as in conventional score fusion. 

The results clearly indicate that non-linear fusion consistently outperforms linear fusion across all settings. Specifically, for conventional score fusion, the min a-DCF decreases from $0.721$ to $0.366$ when moving from linear to non-linear fusion. For MLP-based score fusion with baseline-2~\cite{Shim2022}, non-linear fusion further reduces the min a-DCF from $0.550$ to $0.436$. The proposed Bayesian-optimized MLP-based score fusion architecture achieves the best results, lowering the min a-DCF from $0.335$ to $0.250$, and thereby outperforms the baseline-2 architecture. 

Beyond these improvements, the results consistently show that MLP-based score fusion outperforms conventional score fusion. This highlights the effectiveness of fine-tuning pretrained embedding extractors for their respective tasks (speaker verification or spoof detection), which enhances discriminative power and leads to more effective fusion. Similar performance trends are also observed under EER-based evaluation. Based on these findings, we adopt non-linear fusion of ASV and CM scores using the Bayesian-optimized MLP architecture for all subsequent experiments.

\begin{table}[]
\centering
\caption{Comparison of score fusion and MLP-based classification under linear (LF) and non-linear (NF) fusion. Results are reported in terms of min a-DCF, SV-EER, and SPF-EER. ECAPA-TDNN and AASIST models are used to extract ASV and CM scores, respectively. All results are obtained on the development set.}
\label{tab:score_vs_mlp}
\resizebox{\columnwidth}{!}{%
\begin{tabular}{|cc|c|c|c|c|}
\hline
\multicolumn{2}{|c|}{\textbf{System}} &
  \textbf{Fusion} &
  \textbf{min a-DCF} &
  \textbf{SV-EER (\%)} &
  \textbf{SPF-EER (\%)} \\ \hline
\multicolumn{2}{|c|}{} &
  \cellcolor[HTML]{C0C0C0}LF &
  \cellcolor[HTML]{C0C0C0}0.721 &
  \cellcolor[HTML]{C0C0C0}2.2 &
  \cellcolor[HTML]{C0C0C0}26.8 \\
\multicolumn{2}{|c|}{\multirow{-2}{*}{Score Fusion}} &
  NF &
  \textbf{0.366} &
  1.7 &
  17.8 \\ \hline
\multicolumn{1}{|c|}{} &
   &
  \cellcolor[HTML]{C0C0C0}LF &
  \cellcolor[HTML]{C0C0C0}0.550 &
  \cellcolor[HTML]{C0C0C0}3.3 &
  \cellcolor[HTML]{C0C0C0}26.2 \\
\multicolumn{1}{|c|}{} &
  \multirow{-2}{*}{Baseline} &
  NF &
  \textbf{0.436} &
  3.3 &
  20.7 \\ \cline{2-6} 
\multicolumn{1}{|c|}{} &
   &
  \cellcolor[HTML]{C0C0C0}LF &
  \cellcolor[HTML]{C0C0C0}0.335 &
  \cellcolor[HTML]{C0C0C0}9.6 &
  \cellcolor[HTML]{C0C0C0}14.8 \\
\multicolumn{1}{|c|}{\multirow{-4}{*}{\begin{tabular}[c]{@{}c@{}}MLP-based\\ Classification \\ with\\ Score Fusion\end{tabular}}} &
  \multirow{-2}{*}{Bayesian} &
  NF &
  \textbf{0.250} &
  2.2 &
  11.9 \\ \hline
\end{tabular}%
}
\end{table}

\subsection{Unified SASV with Joint Optimization}

%From the experiments reported in the previous subsection, non-linear fusion of ASV and CM scores were found to outperform conventional linear fusion. 
By adopting the non-linear fusion strategy, we now proceed to comparing %evaluate 
the three systems described in Sec.~\ref{sec:proposed-approach} and illustrated in Fig.~\ref{fig:your_label}. 
Table~\ref{tab:unified_sasv} summarizes the results of joint optimization on ASVspoof5 development set under different configurations, including initialization strategies (random vs. pre-trained MLPs for ASV and CM), loss functions ($\mathcal{L}_\text{sasv}^{(1)}$ in~\eqref{eq:sasv_loss1} vs.\ $\mathcal{L}_\text{sasv}^{(2)}$ in~\eqref{sasv_loss2}) and optimizers (Adam vs. SGD). While random initialization corresponds to the case where ASV and CM MLPs are randomly initialized, pre-trained initialization corresponds to the initializing the ASV and CM MLPs with the weights previously learned for their task with their respective BCE losses.

We first examine joint optimization of ASV and CM MLPs, where trainable MLPs for ASV and CM are jointly optimized using non-linear fusion with the SASV objective (Fig.~\ref{fig:your_label}(a)). Comparing random and pre-trained initialization (the first row in Table~\ref{tab:unified_sasv}) shows negligible differences in performance (min a-DCF $0.321$ vs. $0.318$). While SV-EER slightly increases with pre-trained initialization (2.9\% vs.\ 3.2\%), SPF-EER decreases marginally (15.3\% vs.\ 15.0\%). Given the minimal effect of initialization, we adopt random initialization for subsequent experiments.

We first examine joint optimization of ASV and CM MLPs, where trainable MLPs for ASV and CM are jointly optimized using non-linear fusion with the SASV objective (Fig.~\ref{fig:your_label}(a)). Comparing random and pre-trained initialization (the first row in Table~\ref{tab:unified_sasv}) shows negligible differences in performance (min a-DCF $0.321$ vs. $0.318$). While SV-EER slightly increases with pre-trained initialization (2.9\% vs.\ 3.2\%), SPF-EER decreases marginally (15.3\% vs.\ 15.0\%). Given the minimal effect of initialization, we adopt random initialization for subsequent experiments.

% \textcolor{blue}{
% Additionally, we evaluated the margin-augmented variant of the MLP--MLP configuration under Adam optimization with random initialization, where the cosine embedding loss was incorporated alongside $\mathcal{L}_\text{sasv}^{(1)}$. This configuration yielded a min a-DCF of $0.303$, an SV-EER of $3.1\%$, and an SPF-EER of $14.3\%$. Compared to the baseline $\mathcal{L}_\text{sasv}^{(1)}$ setting ($0.321$ / $2.9\%$ / $15.3\%$), the inclusion of the cosine-based margin improves the overall cost-sensitive metric (min a-DCF) and reduces SPF-EER, indicating enhanced spoof discrimination, at the expense of a marginal increase in SV-EER. 
% These findings suggest that angular margin regularization can beneficially refine the decision boundary for spoof detection while maintaining competitive speaker verification performance.}

Next, we replaced the ASV MLP head with cosine similarity scoring, fusing its calibrated output with the MLP-based CM branch (Fig.~\ref{fig:your_label}(b)). From the results reported in the second row of the Table~\ref{tab:unified_sasv}, this change substantially reduced min a-DCF from $0.321$ to $0.218$ and lowered SPF-EER from 15.3\% to 12.3\%, although SV-EER increased to 4.5\%. We then compared loss functions. Substituting $\mathcal{L}_\text{sasv}^{(1)}$ (a combination of $\mathcal{L}_\text{sasv}^{\text{BCE}}$ and $\mathcal{L}_\text{sasv}^{\text{adcf}}$) with $\mathcal{L}_\text{sasv}^{(2)}$ (a joint aggregation of $\mathcal{L}_\text{sasv}^{\text{adcf}}$, $\mathcal{L}_\text{asv}^{\text{BCE}}$, and $\mathcal{L}_\text{cm}^{\text{BCE}}$) degraded performance, increasing min a-DCF to $0.272$. This indicates that explicitly combining $\mathcal{L}_\text{sasv}^{\text{BCE}}$ with $\mathcal{L}_\text{sasv}^{\text{adcf}}$ is more effective than treating ASV and CM losses independently. Optimizer choice also plays a role: switching from Adam to SGD further reduced min a-DCF from $0.218$ to $0.211$, highlighting the effectiveness of SGD for joint optimization.

%Finally, the proposed unified SASV architecture, which integrates weighted cosine similarity for ASV with an MLP for CM (Fig.~\ref{fig:your_label}(c)), achieved the best overall performance with a min a-DCF of $0.205$ as shown in the last row of the Table~\ref{tab:unified_sasv}. \textcolor{blue}{In the final system, this trend continues, as replacing Adam with SGD further reduced the min a-DCF from $0.282$ to $0.205$.} These results demonstrate that introducing learnable weights at the fusion stage enhances the interaction between ASV and CM subsystems, leading to further gains in SASV performance.

Finally, the proposed unified SASV architecture, which integrates weighted cosine similarity for ASV with an MLP for CM (Fig.~\ref{fig:your_label}(c)), achieved the best overall performance with a min a-DCF of $0.205$ using the SGD optimizer, as shown in the last row of Table~\ref{tab:unified_sasv}. To further examine the impact of the optimization strategy, the same model was also trained with the Adam optimizer, yielding a higher min a-DCF of $0.282$, confirming that SGD remains the more effective choice for this architecture. These findings confirm that introducing learnable weights at the fusion stage enhances the interaction between ASV and CM subsystems, leading to further gains in SASV performance, while margin-based regularization provides complementary benefits primarily in spoof robustness.

\begin{table}[]
\centering
\caption{Results of unified SASV models under joint optimization on the \emph{development set.} 
The \textit{System} columns denote the ASV scoring function and CM network, 
while the \textit{Configuration} columns show the optimizer, loss, and initialization strategy. 
Performance is evaluated using min a-DCF, SV-EER, and SPF-EER, with embeddings from ECAPA-TDNN (ASV) and AASIST (CM).}
\label{tab:unified_sasv}
\resizebox{\columnwidth}{!}{%
\begin{tabular}{|c|c|c|c|c|c|c|c|}
\hline
\multicolumn{2}{|c|}{\textbf{System}} &
\multicolumn{3}{c|}{\textbf{Configuration}} &
\multirow{2}{*}{\textbf{min a-DCF}} &
\multirow{2}{*}{\textbf{SV-EER (\%)}} &
\multirow{2}{*}{\textbf{SPF-EER (\%)}} \\ \cline{1-5}
\textbf{ASV} & \textbf{CM} &
\textbf{Optimizer} &
\textbf{Loss} &
\textbf{Init.} &
 & & \\ \hline

\multirow{2}{*}{MLP} &
\multirow{2}{*}{MLP} &
\multirow{2}{*}{Adam} &
\multirow{2}{*}{$\mathcal{L}_\text{sasv}^{(1)}$} &
Rand. &
0.321 & 2.9 & 15.3 \\ \cline{5-8}
 &  &  &  & \textbf{Pre.} &
0.318 & 3.2 & 15.0 \\ \hline

\multirow{3}{*}{Cosine} &
\multirow{3}{*}{MLP} &
\multirow{2}{*}{Adam} &
$\mathcal{L}_\text{sasv}^{(1)}$ &
\multirow{2}{*}{Rand.} &
\textbf{0.218} & 4.5 & 12.3 \\ \cline{4-4} \cline{6-8}
 &  &  &
\textbf{$\mathcal{L}_\text{sasv}^{(2)}$} &
 & 0.272 & 1.7 & 13.0 \\ \cline{3-8}
 &  &
\textbf{SGD} &
$\mathcal{L}_\text{sasv}^{(1)}$ &
Rand. &
\textbf{0.211} & 2.6 & 11.7 \\ \hline

\multirow{2}{*}{\begin{tabular}[c]{@{}c@{}}Weighted \\ Cosine\end{tabular}} &
\multirow{2}{*}{MLP} &
Adam &
\multirow{2}{*}{$\mathcal{L}_\text{sasv}^{(1)}$} &
\multirow{2}{*}{Rand.} &
0.282 & 1.7 & 13.5 \\ \cline{3-3} \cline{6-8}
 &  &
SGD &  &  &
\textbf{0.205} & 2.8 & 11.1 \\ \hline

\end{tabular}}
\end{table}

\subsection{Evaluation with Various ASV and CM Systems}

The results above used a particular ASV (ECAPA-TDNN) and CM (AASIST) systems. To demonstrate generality of the proposed optimization approach, we now consider more variations in the two 'plug-and-play' subsystems. In particular, Table~\ref{tab:asv_cm_eval} presents a detailed comparison of different ASV–CM pretrained embedding pairings on the evaluation set using the proposed learnable weighted cosine-based unified SASV system. Among the ASV embeddings, ReDimNet performs best, achieving min a-DCF of $0.196$ when paired with SSL-AASIST as CM; and $0.449$ when paired with AASIST as CM. In comparison, WavLM–TDNN achieves $0.215$ and $0.456$, while ECAPA-TDNN achieves $0.204$ and $0.509$. 

These results indicate that integrating SSL-AASIST as the CM consistently improves SASV performance over vanilla AASIST, emphasizing the value of self-supervised embeddings for spoofing detection. Overall, the findings align with trends observed in individual ASV and CM evaluations: ReDimNet~\cite{yakovlev24_interspeech} captures more robust speaker representations than the other models, and SSL-AASIST~\cite{tak22_odyssey} more effectively distinguishes between bonafide and spoofed speech compared to vanilla AASIST.

\begin{table}[]
\centering
\caption{Evaluation of the proposed method with different ASV–CM pairings. 
ASV embeddings are from ECAPA-TDNN, WavLM-TDNN, or ReDimNet, and CM embeddings from AASIST or SSL-AASIST, 
with results reported on the evaluation set.}
\label{tab:asv_cm_eval}
\resizebox{\columnwidth}{!}{%
\begin{tabular}{|c|c|c|c|c|}
\hline
\textbf{ASV}                 & \textbf{CM} & \textbf{min a-DCF} & \textbf{SV-EER (\%)} & \textbf{SPF-EER (\%)} \\ \hline
 & \cellcolor[HTML]{C0C0C0}AASIST & \cellcolor[HTML]{C0C0C0}0.509 & \cellcolor[HTML]{C0C0C0}7.6 & \cellcolor[HTML]{C0C0C0}24.0 \\ \cline{2-5} 
\multirow{-2}{*}{ECAPA-TDNN} & SSL-AASIST  & 0.204    & 8.2                 & 7.8                  \\ \hline
 & \cellcolor[HTML]{C0C0C0}AASIST & \cellcolor[HTML]{C0C0C0}0.456 & \cellcolor[HTML]{C0C0C0}8.9          & \cellcolor[HTML]{C0C0C0}21.1 \\ \cline{2-5} 
\multirow{-2}{*}{WavLM-TDNN} & SSL-AASIST  & 0.215             & 9.8                 & \textbf{7.4}         \\ \hline
 & \cellcolor[HTML]{C0C0C0}AASIST & \cellcolor[HTML]{C0C0C0}0.449     & \cellcolor[HTML]{C0C0C0}\textbf{6.9}            & \cellcolor[HTML]{C0C0C0}21.1    \\ \cline{2-5} 
\multirow{-2}{*}{ReDimNet}   & SSL-AASIST  & \textbf{0.196}  & 8.0 &  7.6  \\ \hline
\end{tabular}%
}
\end{table}

\subsection{Effect of Score Calibration}
\label{sec:scorecalibration}

To quantify the impact of the calibration layers in the best-performing proposed model (Fig.~\ref{fig:your_label}(c)), we evaluate the cost of log-likelihood ratios ($C_{\mathrm{llr}}$) \cite{brummer2006application}, a cross-entropy metric which measures the quality of detection scores interpreted as LLRs.

For this analysis, the ASV and CM scores produced by the proposed model are isolated both \emph{before} and \emph{after} the calibration layers, and the corresponding $C_{\mathrm{llr}}$ values are computed. The ReDimNet and SSL-AASIST models are used to extract ASV and CM embeddings, respectively. The $C_{\mathrm{llr}}$ metric is defined as
%The metric is defined as
\begin{equation}
\mathrm{C}_{\mathrm{llr}} =
\tfrac{1}{2\log 2}\!\bigg(
\frac{1}{|\mathcal{P}|}\sum_{s_i \in \mathcal{P}} \log(1+e^{-s_i})
+\!
\frac{1}{|\mathcal{N}|}\sum_{s_j \in \mathcal{N}} \log(1+e^{s_j})
\bigg)
\end{equation}
where $\mathcal{P}$ and $\mathcal{N}$ denote the sets of positive and negative trial scores, respectively, with $s_i$ and $s_j$ representing the corresponding detection scores. For CM, these correspond to bonafide and spoof trials, while for ASV they correspond to target and non-target trials. Lower $\mathrm{C}_{\mathrm{llr}}$ values indicate better calibrated and more discriminative scores.

As summarized in Table~\ref{tab:cllr_calibration}, calibration reduces $\mathrm{C}_{\mathrm{llr}}$ for both ASV and CM branches. For the ASV branch, cosine similarity scores are naturally bounded and already relatively well calibrated before affine transformation; therefore, calibration provides only a modest improvement in $\mathrm{C}_{\mathrm{llr}}$. In contrast, the CM branch produces raw logits from an MLP classifier with arbitrary scaling, leading to poor calibration when interpreted as LLRs, reflected by the high pre-calibration $\mathrm{C}_{\mathrm{llr}}$. After calibration, $\mathrm{C}_{\mathrm{llr}}$ is substantially reduced, indicating improved score reliability and interpretability. These results highlight the importance of explicit score calibration layers. Since the final decision relies on jointly combining ASV and CM evidence, well-calibrated LLRs ensure that both subsystems contribute on a comparable scale. The reduction in $\mathrm{C}_{\mathrm{llr}}$ therefore supports the effectiveness of the calibration layers and the calibrated integration strategy adopted in this work.

\begin{table}[t]
\centering

\caption{$\mathrm{C}_{\mathrm{llr}}$ scores for the ASV and CM branches before and after score calibration, where lower values indicate better calibration.}
\label{tab:cllr_calibration}
\begin{tabular}{llcc}
\hline
\textbf{Branch} & \textbf{Score Type} & \textbf{Before cal.} & \textbf{After cal.} \\
\hline
ASV & cosine similarity & 0.8553 & \textbf{0.8324} \\
CM & MLP-based logits & 7.0764 & \textbf{0.3687} \\
\hline
\end{tabular}
\end{table}

\subsection{Comparison with Score Fusion Baseline}

\begin{figure}[h]
  \centering
  \includegraphics[width=\columnwidth]{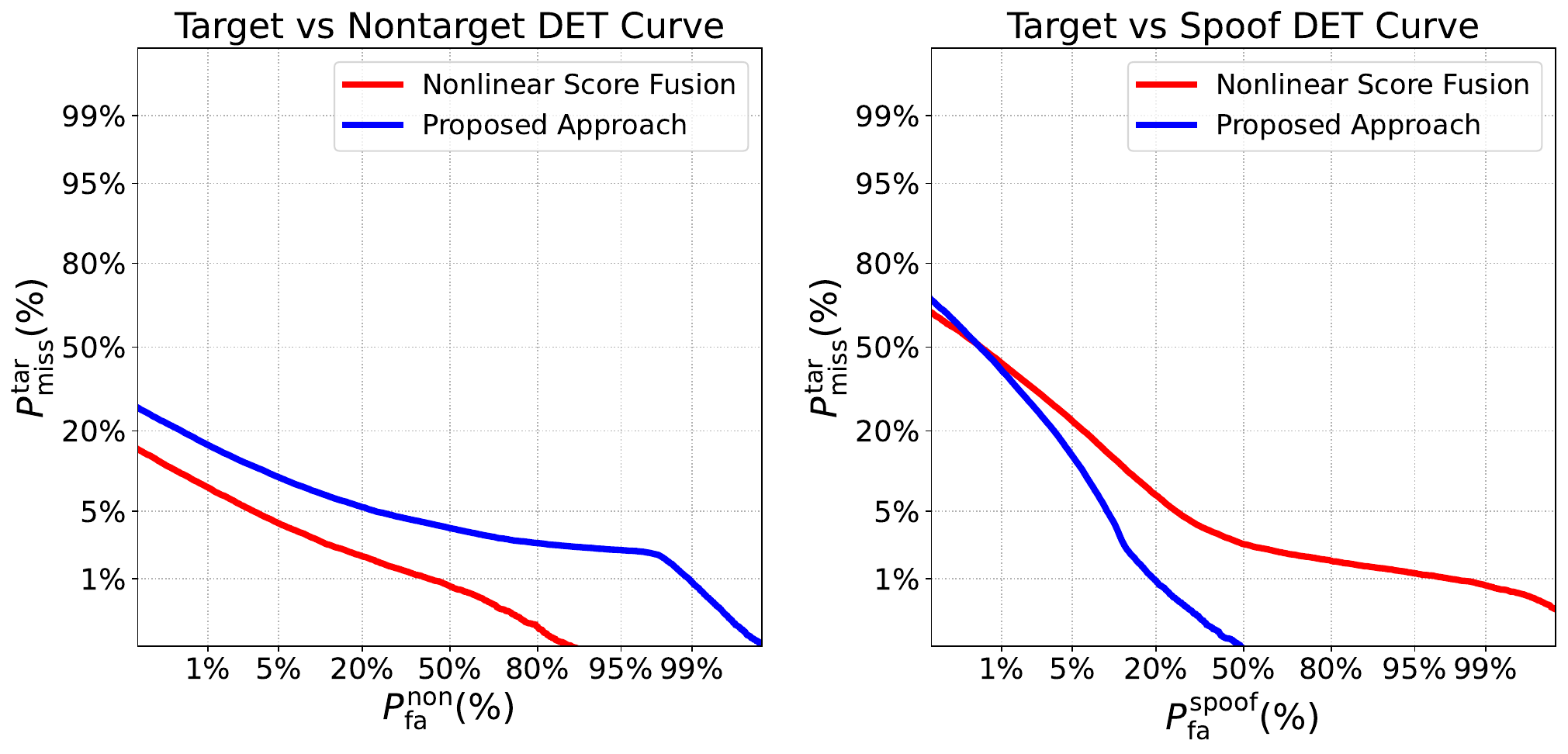}
  \caption{DET curves comparing conventional non-linear score fusion (red) and the proposed approach (blue). The left plot shows the tradeoff between false acceptance of non-target trials ($P^{\text{non.bon}}_{\text{fa}}$) and missed detections of target trials ($P^{\text{tar.bon}}_{\text{miss}}$), corresponding to the conventional ASV performance. The right plot shows the tradeoff between false acceptance of spoof trials ($P^{\text{spf}}_{\text{fa}}$) and missed detections of target trials ($P^{\text{tar}}_{\text{miss}}$), highlighting the system’s spoofing robustness.}
  \label{fig:det_curves}
\end{figure}

We further compare our best unified SASV system against conventional score fusion (as per~\eqref{eq:basic-nonlinear-fusion}) on the evaluation set. Based on Table~\ref{tab:asv_cm_eval}, we use ReDimNet and SSL-AASIST embeddings;  and following Table~\ref{tab:unified_sasv}, the selected SASV architecture integrates cosine-weighted ASV with MLP-based CM. The results, displayed in Table~\ref{tab:final_comparison}, indicate that the proposed system outperforms non-linear score fusion in terms of both minimum and actual DCF. In terms of the two EERs, our system provides substantially increased resilience to spoofing---traded with decrease in target-nontarget speaker discrimination. The DET curves in Fig.~\ref{fig:det_curves} further illustrate this trade-off. 

Such trade-off is expected, since the system needs to balance between potentially conflicting requirements of retaining low miss rate, while providing protection to both non-target speakers and spoofing attacks. Building on our recent study~\cite{kurnaz2024optimizing}, %we note that 
optimization with the combined a-DCF and BCE loss depends on the %prior cost settings and the 
relative weights assigned to $P_{\text{fa}}^{\text{non.bon}}$ and $P_{\text{fa}}^{\text{spf}}$, obtained from the detection costs and class priors. As described in the experimental setup, we fixed the weight of $C_{\text{miss}}^{\text{tar.bon}} \cdot \pi_{\text{tar}}$ to $0.9$, $C_{\text{fa}}^{\text{non.bon}} \cdot \pi_{\text{non}}$ to $0.5$ and $C_{\text{fa}}^{\text{spf}} \cdot \pi_{\text{spf}}$ to $1$. This configuration assigns greater importance to spoofing detection than to speaker verification, %This choice likely 
which explains the improved spoof EER and the higher ASV EER compared to score fusion. Nevertheless, the overall SASV performance (as measured by a-DCF) substantially %under the proposed approach, demonstrating its effectiveness 
outperforms conventional score fusion.

\begin{table}[t!]
\centering
\caption{Final comparison on the evaluation set between the best-performing unified SASV system (ReDimNet + SSL-AASIST with weighted cosine fusion) and conventional non-linear score fusion, where ReDimNet ASV and SSL-AASIST CM scores are combined by non-linear fusion without joint optimization. Actual a-DCF (act a-DCF) values are reported using a uniform threshold derived from the development set ($\tau^{\mathrm{dev}}_{\mathrm{sasv}}$).}
\label{tab:final_comparison}
\resizebox{\columnwidth}{!}{%
\begin{tabular}{|c|c|c|c|c|}
\hline
\textbf{System} & \textbf{min a-DCF} & \textbf{act a-DCF} & \textbf{SV-EER (\%)} & \textbf{SPF-EER (\%)} \\ \hline
Score fusion (SF)    &  0.251  &  0.251  &  \textbf{4.2}  &   11.9  \\ \hline
Proposed (Prop.)       &  \textbf{0.196}   &  \textbf{0.210}   &  8.0  &   \textbf{7.6}  \\ \hline
\end{tabular}%
}
\end{table}

\begin{figure}[h]
  \centering
  \includegraphics[width=\columnwidth]{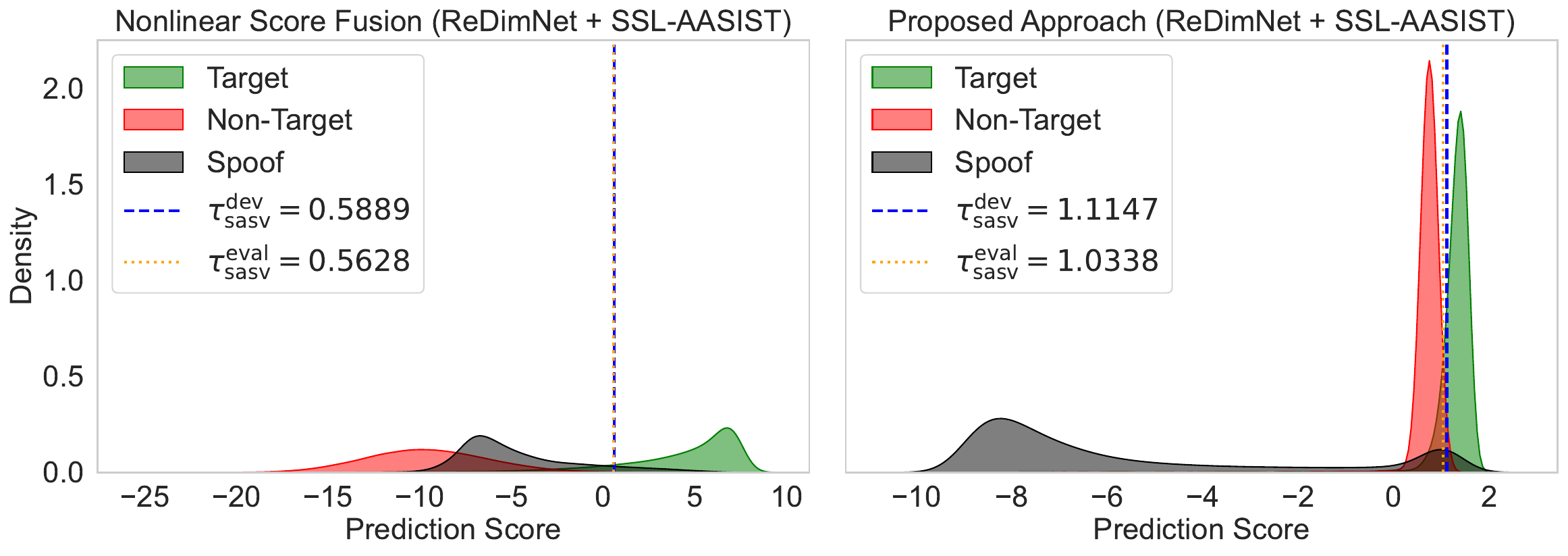}
  \caption{Comparison of non-linear score fusion (left) and the proposed approach (right) using ReDimNet as the ASV system and SSL-AASIST as the CM system. 
The plots show the score distributions for \textit{target}, \textit{non-target}, and \textit{spoof} trials. 
Vertical dashed lines denote the operating thresholds: $\tau^{\text{dev}}_{\text{sasv}}$ (blue dashed line) represents the fixed operating point optimized on the pooled development set and used for the computation of actual metrics, while $\tau^{\text{eval}}_{\text{sasv}}$ (orange dotted line) indicates the theoretical optimal threshold for the evaluation set}.
  \label{fig:ecapa_vs_redimnet}
\end{figure}

%To 
%We further %analyze system behavior, 
%plot 
The score distributions %of the proposed approach and the conventional score fusion 
displayed in Fig.~\ref{fig:ecapa_vs_redimnet} further reveal that the proposed system produces score distributions with lower variance for target, non-target, and spoof trials compared to conventional score fusion. In addition, the proposed system differentiates target speakers more clearly from spoofing attacks, %score distribution than 
than from non-target speakers, %which 
aligned with the above remark %hypothesis based on the current 
about cost and prior settings.

To assess the practical deployment ability of our proposed method, we first computed the empirical threshold on the development set. We then applied this threshold to the evaluation set to compute the actual a-DCF. It is important to note that, in deployment scenarios, we do not have access to evaluation data for threshold estimation; thus, practical use cases must always rely on thresholds derived from the development set. Fig.~\ref{fig:ecapa_vs_redimnet} illustrates the resulting score distributions, and Table~\ref{tab:final_comparison} shows that the proposed system achieves an actual a-DCF of $0.210$, compared to $0.251$ with conventional score fusion.

\subsection{Impact of AAM-Softmax loss}
Margin-based softmax losses have become a standard choice in modern ASV systems due to their ability to force angular margin between speakers, thus enhancing inter-speaker discriminability. Therefore, we investigate the effect of incorporating the ArcFace loss~\cite{deng2019arcface}, also known as additive angular margin (AAM-Softmax) in the speaker verification literature~\cite{9023039, li2022real} within our joint SASV optimization framework. %In the proposed system (Figure~3(c)), 
The best-performing proposed SASV system shown in Fig.~\ref{fig:your_label}(c) %which employs ASV and CM embeddings extracted from RedimNet and SSL-AASIST models, respectively 
is used for this analysis. %we use RedimNet speaker embeddings and SSL-AASIST embeddings as the countermeasure (CM) representation. 
%To investigate the impact of margin-based loss in the proposed joint optimization framework, we %add the 
Concretely, we introduce the additional AAM-Softmax objective to the ASV branch; %of the architecture. 
%Specifically, %we concatenate 
the %ASV 
enrollment and test ASV embeddings are concatenated and passed %them 
to a classifier that determines whether the pair belongs to the same speaker. This auxiliary classification %objective 
task produces the AAM-Softmax loss, which %we combine 
is jointly optimized with the SASV loss ($\mathcal{L}_\text{sasv}^{(1)}$). %during optimization.}

\begin{figure*}[h]
    \centering
    \includegraphics[width=\textwidth]{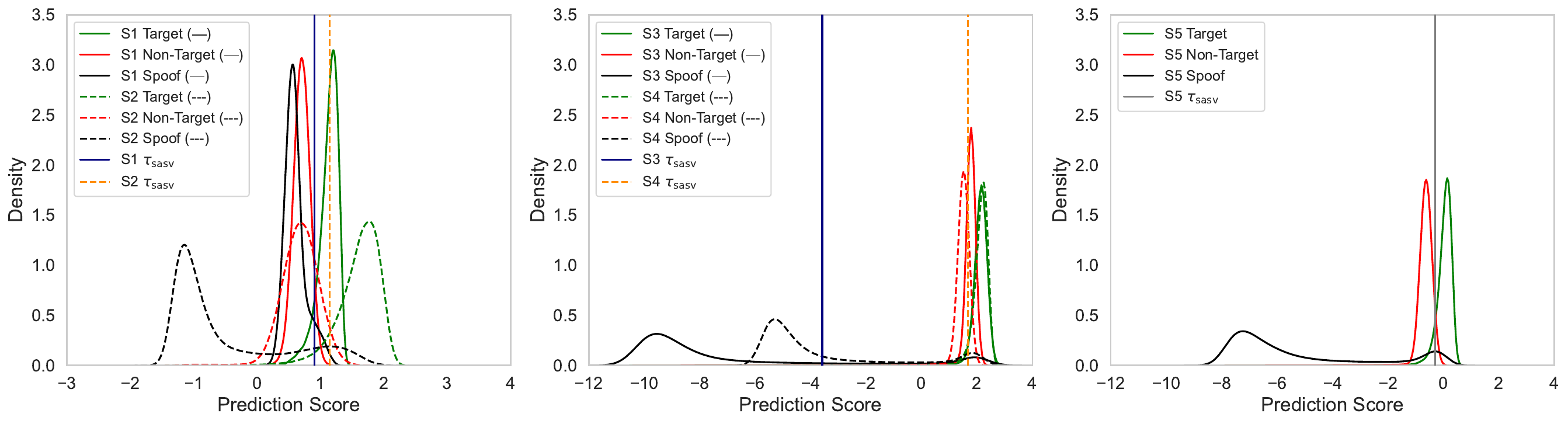}
    \caption{Evaluation-set score distributions of target, non-target, and spoof trials under different operating points. 
    The left and middle panels compare paired operating points (Setting1 (S1) vs Setting2 (S2) and Setting3 (S3) vs Setting4 (S4)), while the right panel shows the distribution for Setting5 (S5).}
    \label{fig:kde_three_panels_settings3_7}
\end{figure*}

\begin{figure}[t]
\centering
\includegraphics[width=\linewidth]{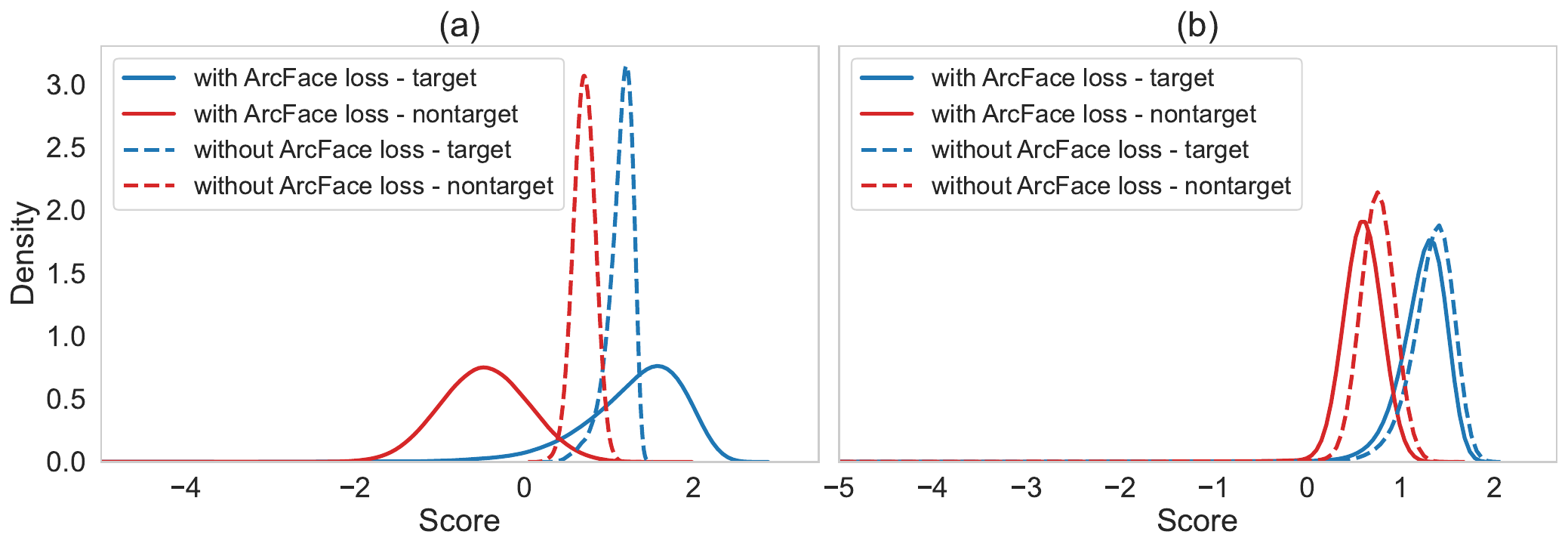}
\caption{Comparison of SASV performance with and without ArcFace loss under two operating points. 
(a) ASV-oriented setting 
($C_{\mathrm{miss}}^{\mathrm{tar.bon}}\!\cdot\!\pi_{\mathrm{tar}}=0.5$, 
$C_{\mathrm{fa}}^{\mathrm{non.bon}}\!\cdot\!\pi_{\mathrm{non}}=0.5$, and 
$C_{\mathrm{fa}}^{\mathrm{spf}}\!\cdot\!\pi_{\mathrm{spf}}=0$.), 
while (b) default setting 
($C_{\mathrm{miss}}^{\mathrm{tar.bon}}\!\cdot\!\pi_{\mathrm{tar}}=0.9$, 
$C_{\mathrm{fa}}^{\mathrm{non.bon}}\!\cdot\!\pi_{\mathrm{non}}=0.5$, and 
$C_{\mathrm{fa}}^{\mathrm{spf}}\!\cdot\!\pi_{\mathrm{spf}}=1.0$).}
\label{fig:arcface_setting_comparison}
\end{figure}

%We evaluate the performance under two configurations: 
To analyze the impact of this margin-based objective under different operating conditions, %the SASV system is evaluated under 
we consider two configurations: (1) an ASV-oriented setting % where 
with the spoofing detection cost set to zero ($C_{\mathrm{miss}}^{\mathrm{tar.bon}} \pi_{\mathrm{tar}} = 0.5,\;
C_{\mathrm{fa}}^{\mathrm{non.bon}} \pi_{\mathrm{non}} = 0.5,\;
C_{\mathrm{fa}}^{\mathrm{spf}} \pi_{\mathrm{spf}} = 0.0)$, and (2) our default setting %used in our experiments 
($C_{\mathrm{miss}}^{\mathrm{tar.bon}} \pi_{\mathrm{tar}} = 0.9,\;
C_{\mathrm{fa}}^{\mathrm{non.bon}} \pi_{\mathrm{non}} = 0.5,\;
C_{\mathrm{fa}}^{\mathrm{spf}} \pi_{\mathrm{spf}} = 1.0$)%used in our experiments
. The obtained results and the corresponding score distributions %of
for ASV target and non-target trials are shown in Table~\ref{tab:ac_comparison} and Figure~\ref{fig:arcface_setting_comparison}, respectively.

Under the ASV-oriented setting, incorporating AAM-Softmax leads to a noticeable improvement in ASV performance: %In particular, 
the min a-DCF %improves 
decreases from $0.145$ to $0.134$, and the SV-EER decreases from $7.9\%$ to $7.1\%$. This improvement is also reflected in the score distributions shown in Figure~\ref{fig:arcface_setting_comparison}(a), where the separation between target and non-target scores becomes more %distinct
pronounced. This demonstrates that the margin-based objective effectively strengthens inter-speaker discrimination when the optimization primarily focuses on ASV-related errors. In contrast, under the default evaluation setting, the %improvement is 
benefit of incorporating AAM-Softmax becomes less consistent. The min a-DCF slightly increases from $0.196$ to $0.207$, while the SV-EER and SPF-EER remain %nearly 
largely unchanged (shown in Figure~\ref{fig:arcface_setting_comparison}(b)). %One 
A possible %reason 
explanation is that the default cost configuration assigns a substantially higher %prior 
weight to %spoofing detection compared to non-target (bonafide) errors,  
spoofing-related errors, which encourages the model to %focus more on separating 
prioritize the separation between bonafide and spoof trials, rather than improving speaker discrimination.

Overall, these findings indicate that AAM-Softmax is particularly beneficial in ASV-dominated operating conditions, where the optimization primarily focuses on speaker discrimination. However, its advantage becomes less pronounced when the objective jointly optimizes for the SASV task.

\begin{table}[t]
\centering

\caption{Performance comparison with and without  ($\mathcal{L}_{\mathrm{AAM}}$) loss under different operating points (OPs).
Operating point A: ASV-oriented setting and B: default setting. }
\label{tab:ac_comparison}
\begin{tabular}{p{0.3cm} p{1.9cm} c c c}
\hline
\textbf{OPs} & \textbf{Loss} & \textbf{min a-DCF} & \textbf{SV-EER (\%)} & \textbf{SPF-EER (\%)} \\
\hline
\multirow{2}{*}{A} 
 & $\mathcal{L}_{\mathrm{sasv}}^{(1)}$ & 0.145 & 7.9 & \textbf{8.2} \\
 & $\mathcal{L}_{\mathrm{sasv}}^{(1)}$ + $\mathcal{L}_{\mathrm{AAM}}$  & \textbf{0.134} & \textbf{7.1} & 10.0 \\
\hline
 \multirow{2}{*}{B} 
 & $\mathcal{L}_{\mathrm{sasv}}^{(1)}$ & \textbf{0.196} & \textbf{8.0} & \textbf{7.6} \\
 & $\mathcal{L}_{\mathrm{sasv}}^{(1)}$ + $\mathcal{L}_{\mathrm{AAM}}$  & 0.207 & 8.1 & 8.1 \\
\hline
\end{tabular}
\end{table}

\subsection{Robustness to Varying Operating Points}

To evaluate the robustness of the proposed approach, five cost configurations in~\eqref{adcf_old} were considered. 
\textbf{Setting1} ($C_{\mathrm{miss}}^{\mathrm{tar.bon}} \cdot \pi_{\mathrm{tar}} = 0.5$, $C_{\mathrm{fa}}^{\mathrm{non.bon}} \cdot \pi_{\mathrm{non}} = 0.5$, $C_{\mathrm{fa}}^{\mathrm{spf}} \cdot \pi_{\mathrm{spf}} = 0$) is an ASV-oriented configuration where spoofing errors incur no cost. 
\textbf{Setting2} ($C_{\mathrm{miss}}^{\mathrm{tar.bon}} \cdot \pi_{\mathrm{tar}} = 0.45$, $C_{\mathrm{fa}}^{\mathrm{non.bon}} \cdot \pi_{\mathrm{non}} = 0.45$, $C_{\mathrm{fa}}^{\mathrm{spf}} \cdot \pi_{\mathrm{spf}} = 0.1$) is an ASV-oriented configuration with a small cost assigned to spoofing errors. 
\textbf{Setting3} ($C_{\mathrm{miss}}^{\mathrm{tar.bon}} \cdot \pi_{\mathrm{tar}} = 0.5$, $C_{\mathrm{fa}}^{\mathrm{non.bon}} \cdot \pi_{\mathrm{non}} = 0$, $C_{\mathrm{fa}}^{\mathrm{spf}} \cdot \pi_{\mathrm{spf}} = 0.5$) is a spoofing-focused configuration where non-target errors incur no cost. 
\textbf{Setting4} ($C_{\mathrm{miss}}^{\mathrm{tar.bon}} \cdot \pi_{\mathrm{tar}} = 0.45$, $C_{\mathrm{fa}}^{\mathrm{non.bon}} \cdot \pi_{\mathrm{non}} = 0.1$, $C_{\mathrm{fa}}^{\mathrm{spf}} \cdot \pi_{\mathrm{spf}} = 0.45$) is a spoofing-focused configuration with a small cost assigned to non-target errors. 
Finally, \textbf{Setting5} ($C_{\mathrm{miss}}^{\mathrm{tar.bon}} \cdot \pi_{\mathrm{tar}} = 1$, $C_{\mathrm{fa}}^{\mathrm{non.bon}} \cdot \pi_{\mathrm{non}} = 1$, $C_{\mathrm{fa}}^{\mathrm{spf}} \cdot \pi_{\mathrm{spf}} = 1$) represents a balanced configuration where all error types have equal cost.

For this analysis, we focus on the best-performing %The 
% proposed 
model (Figure~3(c)). %which %uses 
% employs ASV and CM embeddings extracted from the ReDimNet %speaker embeddings 
% and SSL-AASIST %countermeasure (CM) embeddings
% models, respectively. 
For each setting, %We train five separate models 
a separate SASV model is trained using the soft a-DCF objective %function, each corresponding to one of the above settings
with the corresponding prior-cost weights. Each trained model is evaluated using the aDCF metric under all five evaluation settings, with threshold optimization following our earlier study in~\cite[Algorithm1]{kurnaz2024optimizing}. Since the operating conditions vary across settings, each condition requires a different, optimized decision threshold. %This enables analysis of both matched and mismatched training-evaluation conditions with respect to the prior-cost configurations. 

The %evaluation 
results are summarized in Table~\ref{tab:op_points}. %We observe that 
For Setting1 and Setting3, which strongly emphasize either ASV or spoofing detection, the best performance is %achieved 
obtained with the same configuration being used for evaluation. This indicates that the model effectively adapts to the dominant task emphasized during training. For Setting2 and Setting4, where a small cost ($0.1$) is introduced for spoofing detection or non-target errors, the minimum a-DCF %does not occur 
is not observed exactly at the corresponding evaluation setting. Nonetheless, the best performance %appears 
occurs at the %nearest 
closest neighboring %setting 
configuration (Setting1 for Setting2 and Setting3 for Setting4). This suggests that the model %behavior smoothly 
transitions smoothly between closely related prior-cost configurations. For the balanced configuration (Setting5), where equal importance is assigned to all three error types, the lowest a-DCF is %observed at 
obtained under Setting1, followed by Setting3, rather than %exactly at 
under Setting5 itself. This indicates that the model tends to favor configurations where one task is slightly %more 
emphasized. %which may %lead to improved 
%improve score separability.}

Similar trends appear in the score distributions in Fig.~\ref{fig:kde_three_panels_settings3_7}. In Settings 1 and 2, where ASV is emphasized, target and non-target scores show clearer separation. Compared to Setting1, Setting2 yields slightly improved spoof score separation due to the small spoofing cost. A similar pattern occurs between Settings 3 and 4, where the emphasis shifts toward spoofing detection. Under the balanced configuration, non-target and spoof scores exhibit comparable separation, reflecting equal importance assigned to both tasks.

Overall, these observations provide useful insights into how different prior-cost configurations influence model behavior. While some inconsistencies appear in the evaluation, they %highlight 
%reveal interesting characteristics of the proposed approach and %suggest opportunities 
highlight potential directions for future work.

\begin{table}[t]
\centering
\caption{Cross-evaluation matrix (min a-DCF) of different operating points on the evaluation set. Rows denote the training operating point, and columns denote the evaluation operating point.}
\label{tab:op_points}
\begin{tabular}{c c c c c c}
\hline
\textbf{Train $\backslash$ Eval} & \textbf{Setting1} & \textbf{Setting2} & \textbf{Setting3} & \textbf{Setting4} & \textbf{Setting5} \\
\hline
\textbf{Setting1} & \textbf{0.145} & 0.160 & 0.161 & 0.178 & 0.207 \\
\textbf{Setting2} & \textbf{0.136} & 0.155 & 0.173 & 0.190 & 0.217 \\
\textbf{Setting3} & 0.180 & 0.190 & \textbf{0.132} & 0.183 & 0.223 \\
\textbf{Setting4} & 0.151 & 0.164 & \textbf{0.141} & 0.173 & 0.207 \\
\textbf{Setting5} & \textbf{0.144} & 0.159 & 0.150 & 0.176 & 0.208 \\
\hline
\end{tabular}
\end{table}

\begin{table*}[t]
\footnotesize
\centering
\caption{\footnotesize
Attack-wise comparison between conventional non-linear score fusion (SF) and different operating points (Default Setting, Setting2, and Setting4) configuration on the evaluation set. Actual a-DCF (act a-DCF) values are reported using a uniform threshold derived from the development set ($\tau^{\mathrm{dev}}_{\mathrm{sasv}}$). 
Performance is reported in terms of SPF-EER (\%), min a-DCF, and act a-DCF, grouped by vocoder type. 
Row colors indicate vocoder type: \textcolor{hifiganbg}{\rule{0.3cm}{0.3cm}} HiFi-GAN, 
\textcolor{wavconcatbg}{\rule{0.3cm}{0.3cm}} Wav. Concat., 
\textcolor{bigvganbg}{\rule{0.3cm}{0.3cm}} BigVGAN, 
\textcolor{advattbg}{\rule{0.3cm}{0.3cm}} Adv. Att.}

\label{tab:attack_wise_comparison_grouped}

\resizebox{\textwidth}{!}{
\begin{tabular}{|c|c|ccc|ccc|ccc|ccc|}
\hline
\multirow{3}{*}{\textbf{ID}} & \multirow{3}{*}{\textbf{Category}}
& \multicolumn{3}{c|}{\multirow{2}{*}{\textbf{SF}}}
& \multicolumn{9}{c|}{\textbf{Proposed}} \\
\cline{6-14}
& & & & &
\multicolumn{3}{c|}{\textbf{Default Setting}}
& \multicolumn{3}{c|}{\textbf{Setting2}}
& \multicolumn{3}{c|}{\textbf{Setting4}} \\
\cline{3-14}
& & \textbf{SPF-EER} & \textbf{min a-DCF} & \textbf{act a-DCF}
  & \textbf{SPF-EER} & \textbf{min a-DCF} & \textbf{act a-DCF}
  & \textbf{SPF-EER} & \textbf{min a-DCF} & \textbf{act a-DCF}
  & \textbf{SPF-EER} & \textbf{min a-DCF} & \textbf{act a-DCF} \\
\hline

\rowcolor{hifiganbg} A17 & Zero-shot TTS & 7.5 & 0.155 & 0.173 & 2.0 & 0.154 & 0.177 & 2.9 & 0.139 & 0.162 & 1.8 & 0.107 & 0.132 \\
\rowcolor{hifiganbg} A24 & Zero-shot VC  & 10.6 & 0.209 & 0.210 & 9.1 & 0.186 & 0.196 & 9.0 & 0.153 & 0.170 & 8.6 & 0.182 & 0.183 \\
\rowcolor{hifiganbg} A25 & Zero-shot VC  & 3.6 & 0.083 & 0.143 & 1.9 & 0.150 & 0.173 & 1.9 & 0.136 & 0.160 & 1.9 & 0.101 & 0.127 \\
\rowcolor{hifiganbg} A26 & Zero-shot VC  & 4.9 & 0.101 & 0.145 & 2.5 & 0.151 & 0.174 & 2.6 & 0.137 & 0.160 & 2.4 & 0.104 & 0.128 \\
\rowcolor{hifiganbg} A28 & Zero-shot TTS & 23.5 & 0.467 & 0.620 & 24.2 & 0.368 & 0.370 & 23.4 & 0.234 & 0.240 & 22.3 & 0.446 & 0.500 \\
\rowcolor{hifiganbg} A29 & Zero-shot TTS & 4.6 & 0.100 & 0.147 & 1.1 & 0.150 & 0.174 & 1.3 & 0.137 & 0.160 & 1.1 & 0.101 & 0.128 \\
\hline

\rowcolor{wavconcatbg} A19 & Few-shot TTS & 10.5 & 0.218 & 0.219 & 3.5 & 0.162 & 0.184 & 4.9 & 0.144 & 0.166 & 2.6 & 0.114 & 0.138 \\
\hline

\rowcolor{bigvganbg} A21 & Zero-shot TTS & 5.1 & 0.103 & 0.145 & 1.0 & 0.149 & 0.173 & 1.2 & 0.136 & 0.160 & 1.0 & 0.100 & 0.126 \\
\rowcolor{bigvganbg} A22 & Zero-shot TTS & 5.7 & 0.122 & 0.151 & 1.9 & 0.151 & 0.174 & 2.3 & 0.137 & 0.160 & 1.8 & 0.102 & 0.128 \\
\hline

\rowcolor{advattbg} A18 & Malafide  & 15.1 & 0.308 & 0.323 & 11.1 & 0.209 & 0.219 & 11.4 & 0.163 & 0.181 & 8.8 & 0.192 & 0.198 \\
\rowcolor{advattbg} A20 & Malafide  & 10.6 & 0.221 & 0.222 & 5.1 & 0.167 & 0.186 & 6.1 & 0.145 & 0.167 & 3.9 & 0.123 & 0.142 \\
\rowcolor{advattbg} A23 & Malafide  & 11.5 & 0.230 & 0.231 & 7.3 & 0.177 & 0.191 & 7.6 & 0.149 & 0.168 & 6.0 & 0.148 & 0.159 \\
\rowcolor{advattbg} A27 & Malacopula & 13.8 & 0.283 & 0.290 & 10.5 & 0.203 & 0.214 & 10.7 & 0.161 & 0.178 & 8.5 & 0.184 & 0.188 \\
\rowcolor{advattbg} A30 & Malafide+Malacopulo & 20.3 & 0.424 & 0.457 & 16.7 & 0.262 & 0.269 & 17.0 & 0.187 & 0.202 & 13.9 & 0.273 & 0.276 \\
\rowcolor{advattbg} A31 & Malacopula & 16.5 & 0.348 & 0.353 & 12.7 & 0.223 & 0.235 & 13.1 & 0.170 & 0.187 & 10.4 & 0.216 & 0.223 \\
\rowcolor{advattbg} A32 & Malacopula & 8.7 & 0.180 & 0.188 & 4.6 & 0.161 & 0.181 & 5.0 & 0.142 & 0.163 & 3.6 & 0.117 & 0.137 \\
\hline
\end{tabular}}
\end{table*}

\subsection{Attack-wise Robustness and Operating Point Analysis}

As a final analysis, Table~\ref{tab:attack_wise_comparison_grouped} presents the attack-wise breakdown of SPF-EER (\%), min a-DCF, and actual a-DCF for the proposed unified SASV system and the conventional score fusion (SF) baseline across these operating points. The Setting2 configuration, which prioritizes spoof rejection, achieves the lowest SPF-EER for most attacks, confirming its specialization. For example, compared to the SF baseline, Setting2 reduces the SPF-EER from $10.53\%$ to $2.65\%$ for A19 and from $5.15\%$ to $1.01\%$ for A21. Improvements are also observed in both min and actual a-DCF for most attacks, indicating that the benefits of joint optimization extend to operational cost-based evaluations.

The intermediate configurations Default Setting and Setting4 maintain lower SPF-EER than the SF baseline in most cases while providing a more balanced trade-off between spoof detection and speaker verification. When analyzed by vocoder families, the proposed method shows clear gains for BigVGAN-based attacks (A25--A27). In HiFi-GAN-based cases (A28--A30), spoof detection remains strong, although some VC and TTS systems exhibit slightly higher a-DCF values compared to the most specialized configuration.

Across all configurations, adversarial and high-quality attacks (A18--A32) remain the most challenging conditions in ASVspoof 5. Nevertheless, the proposed approach achieves its largest relative gains in these scenarios. For instance, for attack A30, Setting4 reduces the min a-DCF from $0.424$ to $0.303$ and the SPF-EER from $20.32\%$ to $10.61\%$ compared to the SF baseline. Similar improvements are observed for A31, where Setting4 reduces SPF-EER from $16.51\%$ to $8.41\%$. Overall, consistent trends across both min and actual a-DCF confirm the robustness of the proposed joint optimization framework.

\subsection{Comparison with Top-Perfoming Systems on ASVspoof 5 Challenge}

It is instructive to compare our results with prior studies. Table~\ref{tab:asvspoof5_top_comparison} contrasts our best-performing system (ReDimNet + SSL-AASIST with joint non-linear optimization) with the top-ranked systems from the \textbf{ASVspoof 5 Track 2 (Open Condition)}, including T45~\cite{chen24_asvspoof}, T36~\cite{okhotnikov24_asvspoof}, and T39~\cite{aliyev24_asvspoof}. These high-performing systems rely on extensive calibration, data engineering, and large ensembles (e.g., T45 uses 12 subsystems and T36 fuses 6 CM models). In contrast, our model achieves a min a-DCF of $0.196$ using a single ASV–CM backbone pair and without extensive data augmentation, highlighting the simplicity of the proposed pipeline and showing that jointly optimized non-linear score fusion can provide a practical solution with a lightweight setup.

\begin{table}[h]
\centering
\caption{Comparison with top-performing systems on ASVspoof 5 Track 2 (Open Condition). Results are reported in terms of pooled min a-DCF on the evaluation set.}
\label{tab:asvspoof5_top_comparison}
\begin{tabular}{llc}
\hline
\textbf{Team ID} & \textbf{Approach} & \textbf{min a-DCF} \\
\hline
T45 \cite{chen24_asvspoof} & Ensemble (12 systems) & \textbf{0.0756} \\
T36 \cite{okhotnikov24_asvspoof} & Ensemble (6 CM + 1 ASV) & 0.1156 \\
T39 \cite{aliyev24_asvspoof} & Ensemble (ResNet + WavLM based models) & 0.1203 \\
\hline
\textbf{Ours} & \textbf{ReDimNet + SSL-AASIST} & 0.1960 \\
\hline
\end{tabular}
\end{table}

\section{Conclusion}

In this work, we addressed the \textbf{spoofing-robust automatic speaker verification} (SASV) task within a unified, modular framework. Our proposed system integrates learnable weighted cosine scoring for ASV with an MLP-based CM backend, jointly optimized using task-specific and cost-sensitive losses. Extensive experiments show that the unified SASV system achieves consistent improvements over conventional score fusion under the specified decision costs and priors across min a-DCF, actual a-DCF, and EER metrics.

Our analysis reveals several key findings. First, as expected, self-supervised embeddings improve robustness: ReDimNet provides stronger speaker representations for ASV, while SSL-AASIST enhances spoof–bonafide discrimination compared to vanilla AASIST. Second, the joint use of cross-entropy and a-DCF losses outperforms other evaluated loss combinations. Third, \textbf{weighted cosine scoring for the ASV branch} proves particularly effective, better aligning with the verification task viewed as a detection problem and improving discriminability over a purely trainable projection head. Finally, the proposed system shows robustness against diverse spoofing attacks, including adversarial scenarios, consistently outperforming conventional score fusion.

A limitation of the proposed approach is that a-DCF optimization requires predefined class priors and detection costs, typically selected heuristically for a single operating point. Our experiments show that changing these settings between training and evaluation may introduce inconsistencies, where the minimum a-DCF occurs near, but not always exactly at, the intended operating point. Nevertheless, the results highlight the effectiveness of jointly optimized, embedding-based unified SASV architectures and the benefits of cost-aware objectives. Future work will explore optimizing performance across a \emph{range} of operating points, adaptive cost weighting, and improved cross-dataset generalization to enhance the practical deployment of spoofing-robust speaker verification systems.

\bibliographystyle{IEEEtran}
\bibliography{bibliography}

@inproceedings{muller25_interspeech,
  title     = {{Replay Attacks Against Audio Deepfake Detection}},
  author    = {Nicolas Müller and et al.},
  booktitle = {Proc. {Interspeech 2025}},
  pages     = {2245--2249},
  doi       = {10.21437/Interspeech.2025-20},
  issn      = {2958-1796},
}

@inproceedings{leeuwen13_interspeech,
  title     = {The distribution of calibrated likelihood-ratios in speaker recognition},
  author    = {David A. van Leeuwen and Niko Brümmer},
  booktitle = {Proc. Interspeech 2013},
  pages     = {1619--1623},
  doi       = {10.21437/Interspeech.2013-406},
  issn      = {2958-1796},
}

@inproceedings{das20c_interspeech,
  title     = {The Attacker’s Perspective on Automatic Speaker Verification: An Overview},
  author    = {Rohan Kumar Das and Xiaohai Tian and Tomi Kinnunen and Haizhou Li},
  booktitle = {Proc. Interspeech 2020},
  pages     = {4213--4217},
  doi       = {10.21437/Interspeech.2020-1052},
  issn      = {2958-1796},
}

@ARTICLE{sizov2015,
  author={Sizov, Aleksandr and Khoury, Elie and Kinnunen, Tomi and Wu, Zhizheng and Marcel, Sébastien},
  journal={IEEE Transactions on Information Forensics and Security}, 
  title={Joint Speaker Verification and Antispoofing in the  $i$ -Vector Space}, 
  year={2015},
  volume={10},
  number={4},
  pages={821-832},
  doi={10.1109/TIFS.2015.2407362}
}

@ARTICLE{tdcf_kinnunen2020,
  author={Kinnunen, Tomi and et al.},
  journal={IEEE/ACM Transactions on Audio, Speech, and Language Processing}, 
  title={Tandem Assessment of Spoofing Countermeasures and Automatic Speaker Verification: Fundamentals}, 
  year={2020},
  volume={28},
  number={},
  pages={2195-2210},
  doi={10.1109/TASLP.2020.3009494}
}

@inproceedings{todisco18_interspeech,
  title     = {Integrated Presentation Attack Detection and Automatic Speaker Verification: Common Features and Gaussian Back-end Fusion},
  author    = {Massimiliano Todisco and Héctor Delgado and Kong Aik Lee and Md Sahidullah and Nicholas Evans and Tomi Kinnunen and Junichi Yamagishi},
  booktitle = {Proc. Interspeech 2018},
  pages     = {77--81},
  doi       = {10.21437/Interspeech.2018-2289},
  issn      = {2958-1796},
}

@book{duda2000pattern,
  author    = {Richard O. Duda and Peter E. Hart and David G. Stork},
  title     = {Pattern Classification},
  edition   = {2},
  publisher = {John Wiley \& Sons},
  address   = {New York},
  year      = {2001},
  isbn      = {978-0471056690}
}

@inproceedings{asali25_interspeech,
  title     = {{ATMM-SAGA: Alternating Training for Multi-Module with Score-Aware Gated Attention SASV system}},
  author    = {Amro Asali and Yehuda Ben-Shimol and Itshak Lapidot},
  booktitle = {{Proc. Interspeech 2025}},
  pages     = {3708--3712},
  doi       = {10.21437/Interspeech.2025-529},
  issn      = {2958-1796},
}

@inproceedings{li25h_interspeech,
  title     = {{Bayesian Learning for Domain-Invariant Speaker Verification and Anti-Spoofing}},
  author    = {Jin Li and Man-Wai Mak and Johan Rohdin and Kong Aik Lee and Hynek Hermansky},
  booktitle = {{Proc. Interspeech 2025}},
  pages     = {1123--1127},
  doi       = {10.21437/Interspeech.2025-655},
  issn      = {2958-1796},
}

@inproceedings{akiba2019optuna,
  title={Optuna: A next-generation hyperparameter optimization framework},
  author={Akiba, Takuya and Sano, Shotaro and Yanase, Toshihiko and Ohta, Takeru and Koyama, Masanori},
  booktitle={Proc. ACM SIGKDD 2019},
  pages={2623--2631},
}

@inproceedings{tak22_odyssey,
  title     = {Automatic Speaker Verification Spoofing and Deepfake Detection Using Wav2vec 2.0 and Data Augmentation},
  author    = {Hemlata Tak and Massimiliano Todisco and Xin Wang and Jee-weon Jung and Junichi Yamagishi and Nicholas Evans},
  booktitle = {Proc. Odyssey 2022},
  pages     = {112--119},
  doi       = {10.21437/Odyssey.2022-16},
}

@ARTICLE{9814838,
  author={Chen, Sanyuan and et al.},
  journal={IEEE Journal of Selected Topics in Signal Processing}, 
  title={WavLM: Large-Scale Self-Supervised Pre-Training for Full Stack Speech Processing}, 
  year={2022},
  volume={16},
  number={6},
  pages={1505-1518},
  doi={10.1109/JSTSP.2022.3188113}}

@inproceedings{buker25_interspeech,
  title     = {{Evaluating Parameter Sharing for Spoofing-Aware Speaker Verification: A Case Study on the ASVspoof 5 Dataset}},
  author    = {Aykut Büker and Oğuzhan Kurnaz and Şule Bekiryazıcı and Selim Can Demirtaş and Cemal Hanilçi},
  booktitle = {{Proc. Interspeech 2025}},
  pages     = {4573--4577},
  doi       = {10.21437/Interspeech.2025-2618},
  issn      = {2958-1796},
}

@book{Jaynes03-prob-theory,
  address = {Cambridge},
  author = {Jaynes, E. T.},
  publisher = {Cambridge University Press},
  title = {Probability theory: The logic of science},
  year = 2003
}

@ARTICLE{Kinnunen2024-tEER,
  author={Kinnunen, Tomi H. and Lee, Kong Aik and Tak, Hemlata and Evans, Nicholas and Nautsch, Andreas},
  journal={IEEE Transactions on Pattern Analysis and Machine Intelligence}, 
  title={t-EER: Parameter-Free Tandem Evaluation of Countermeasures and Biometric Comparators}, 
  year={2024},
  volume={46},
  number={5},
  pages={2622-2637},
  doi={10.1109/TPAMI.2023.3313648}}

@inproceedings{li2022real,
  title={Real additive margin softmax for speaker verification},
  author={Li, Lantian and Nai, Ruiqian and Wang, Dong},
  booktitle={in Proc. ICASSP 2022},
  pages={7527--7531},
  organization={IEEE}
}

@INPROCEEDINGS{9023039,
  author={Xiang, Xu and Wang, Shuai and Huang, Houjun and Qian, Yanmin and Yu, Kai},
  booktitle={in Proc. APSIPA ASC 2019}, 
  title={Margin Matters: Towards More Discriminative Deep Neural Network Embeddings for Speaker Recognition}, 
  year={2019},
  volume={},
  number={},
  pages={1652-1656},
  doi={10.1109/APSIPAASC47483.2019.9023039}}

@inproceedings{deng2019arcface,
  title={Arcface: Additive angular margin loss for deep face recognition},
  author={Deng, Jiankang and Guo, Jia and Xue, Niannan and Zafeiriou, Stefanos},
  booktitle={in Proc. CVPR 2019},
  pages={4690--4699},
}

@inproceedings{Zhang2022,
  author    = {You Zhang and Ge Zhu and Zhiyao Duan},
  title     = {{A Probabilistic Fusion Framework for Spoofing Aware Speaker Verification}},
  booktitle = {Proc. Odyssey 2022},
  pages     = {77--84},
  doi       = {10.21437/Odyssey.2022-11}
}

@article{Doddington2000-NIST,
title = {The {NIST} speaker recognition evaluation – Overview, methodology, systems, results, perspective},
author = {George R. Doddington and Mark A. Przybocki and Alvin F. Martin and Douglas A. Reynolds},
journal = {Speech Communication},
volume = {31},
number = {2},
pages = {225-254},
year = {2000},
issn = {0167-6393}
}

@inproceedings{shim2024adcf,
  title     = {{a-DCF}: an architecture agnostic metric with application to spoofing-robust speaker verification},
  author    = {Hye-jin Shim and Jee-weon Jung and Tomi Kinnunen and Nicholas Evans and Jean-François Bonastre and Itshak Lapidot},
  booktitle = {Proc. Odyssey 2024},
  pages     = {158--164},
  doi       = {10.21437/odyssey.2024-23},
}

@inproceedings{wang24l_interspeech,
  title     = {Revisiting and Improving Scoring Fusion for Spoofing-aware Speaker Verification Using Compositional Data Analysis},
  author    = {Xin Wang and Tomi Kinnunen and Kong Aik Lee and Paul-Gauthier Noé and Junichi Yamagishi},
  booktitle = {Proc. Interspeech 2024},
  pages     = {1110--1114},
  doi       = {10.21437/Interspeech.2024-422},
  issn      = {2958-1796},
}

@inproceedings{Shim2022,
  title={Baseline Systems for the First Spoofing-Aware Speaker Verification Challenge: Score and Embedding Fusion},
  author={Shim, Hye-Jin and Tak, Hemlata and Liu, Xuechen and Heo, Hee-Soo and Jung, Jee-Weon and Chung, Joon Son and Chung, Soo-Whan and Yu, Ha-Jin and Lee, Bong-Jin and Todisco, Massimiliano and others},
  booktitle={Proc. Odyssey 2022},
}

@inproceedings{Lee2022,
  title     = {{Representation Selective Self-distillation and wav2vec 2.0 Feature Exploration for Spoof-aware Speaker Verification}},
  author    = {{Jin Woo Lee and Eungbeom Kim and Junghyun Koo and Kyogu Lee}},
  booktitle = {{Proc. Interspeech 2022}},
  pages     = {{2898--2902}},
  doi       = {{10.21437/Interspeech.2022-11460}},
  issn      = {{2958-1796}},
}

@article{Jung2022,
   author = {Jee-weon Jung and et al.},
   pages = {1-8},
   title = {{SASV} Challenge 2022: A Spoofing Aware Speaker Verification Challenge Evaluation Plan},
   url = {http://arxiv.org/abs/2201.10283},
   year = {2022},
}

@inproceedings{Wang2022b,
  author    = {Xingming Wang and Xiaoyi Qin and Yikang Wang and Yunfei Xu and Ming Li},
  title     = {{The DKU-OPPO System for the 2022 Spoofing-Aware Speaker Verification Challenge}},
  year      = {2022},
  booktitle = {Proc. Interspeech 2022},
  pages     = {4396--4400},
  doi       = {10.21437/Interspeech.2022-11190},
  issn      = {2958-1796}
}

@inproceedings{teng22_interspeech,
  title     = {{SA-SASV}: An End-to-End Spoof-Aggregated Spoofing-Aware Speaker Verification System},
  author    = {Zhongwei Teng and Quchen Fu and Jules White and Maria Powell and Douglas Schmidt},
  booktitle = {Proc. Interspeech 2022},
  pages     = {4391--4395},
  doi       = {10.21437/Interspeech.2022-11029},
  issn      = {2958-1796},
}

@inproceedings{Alenin2022,
  title     = {{A Subnetwork Approach for Spoofing Aware Speaker Verification}},
  author    = {{Alexander Alenin and Nikita Torgashov and Anton Okhotnikov and Rostislav Makarov and Ivan Yakovlev}},
  booktitle = {{Proc. Interspeech 2022}},
  pages     = {{2888--2892}},
  doi       = {{10.21437/Interspeech.2022-10921}},
  issn      = {{2958-1796}},
}

@inproceedings{sutskever2013importance,
  title={On the importance of initialization and momentum in deep learning},
  author={Sutskever, Ilya and Martens, James and Dahl, George and Hinton, Geoffrey},
  booktitle={Proc. ICML 2013},
  pages={1139--1147},
}

@inproceedings{Zhang2022sasv,
  title={{SASV} Based on Pre-trained {ASV} System and Integrated Scoring Module},
  author={Zhang, Yuxiang and Li, Zhuo and Wang, Wenchao and Zhang, Pengyuan},
  booktitle={Proc. Interspeech 2022},
  pages={4376--4380},
}

@inproceedings{brummer13_interspeech,
  title     = {Likelihood-ratio calibration using prior-weighted proper scoring rules},
  author    = {Niko Brümmer and George R. Doddington},
  year      = {2013},
  booktitle = {Interspeech 2013},
  pages     = {1976--1980},
  doi       = {10.21437/Interspeech.2013-470},
  issn      = {2958-1796},
}

@inproceedings{ferrer20b_odyssey,
  title     = {A Speaker Verification Backend for Improved Calibration Performance across Varying Conditions},
  author    = {Luciana Ferrer and Mitchell McLaren},
  year      = {2020},
  booktitle = {The Speaker and Language Recognition Workshop (Odyssey 2020)},
  pages     = {372--379},
  doi       = {10.21437/Odyssey.2020-52},
}

@inproceedings{mun23_interspeech,
  title     = {Towards Single Integrated Spoofing-aware Speaker Verification Embeddings},
  author    = {Sung Hwan Mun and et al.},
  booktitle = {Proc. Interspeech 2023},
  pages     = {3989--3993},
  doi       = {10.21437/Interspeech.2023-1402},
  issn      = {2958-1796},
}

@inproceedings{kang22_interspeech,
  author={Woohyun Kang and Md Jahangir Alam and Abderrahim Fathan},
  title={{End-to-end framework for spoof-aware speaker verification}},
  booktitle={Proc. Interspeech 2022},
  pages={4362--4366},
  doi={10.21437/Interspeech.2022-139}
}

@techreport{ISOpresentationAtack,
  type = {{Standard}},
  address = {Geneva, Switzerland},
  key = {{ISO/IEC 30107}},
  year = {2016},
  title = {{ISO/IEC 30107. Information Technology -- Biometric presentation attack detection}},
  institution = {International Organization for Standardization}
}

@article{brummer2006application,
  title={Application-independent evaluation of speaker detection},
  author={Br{\"u}mmer, Niko and Du Preez, Johan},
  journal={Computer Speech \& Language},
  volume={20},
  number={2-3},
  pages={230--275},
  year={2006},
  publisher={Elsevier}
}

@inproceedings{mingote19_interspeech,
  author    = {Victoria Mingote and et al.},
  title     = {{Optimization of False Acceptance/Rejection Rates and Decision Threshold for End-to-End Text-Dependent Speaker Verification Systems}},
  booktitle = {Proc. Interspeech 2019},
  pages     = {2903--2907},
  doi       = {10.21437/Interspeech.2019-2550},
  issn      = {2308-457X}
}

@inproceedings{reynolds94_asriv,
  author={Douglas A. Reynolds},
  title={{Speaker identification and verification using Gaussian mixture speaker models}},
  year=1994,
  booktitle={Proc. ESCA Workshop on Automatic Speaker Recognition, Identification and Verification},
  pages={27--30}
}

@article{reynolds1995,
    author = {Reynolds, D. A. and Rose, R. C.},
    title = {Robust Text-Independent Speaker Identification Using Gaussian Mixture Speaker Models},
    journal = {IEEE Transactions on Speech and Audio Processing},
    volume = {3},
    number = {1},
    pages = {72-83},
    year = {1995},
}

@article{Morrison2013-calibration,
  author  = {Geoffrey Stewart Morrison},
  title   = {Tutorial on logistic-regression calibration and fusion: Converting a score to a likelihood ratio},
  journal = {Australian Journal of Forensic Sciences},
  year    = {2013},
  volume  = {45},
  number  = {2},
  pages   = {173--197},
  doi     = {10.1080/00450618.2012.733025}
}

@phdthesis{Brummer2010,
  author       = {Niko Br{\"u}mmer},
  title        = {Measuring, refining and calibrating speaker and language information extracted from speech},
  school       = {University of Stellenbosch},
  year         = {2010},
  month        = {December},
  type         = {Ph.D. dissertation}
}

@article{RAMOS2013-reliable-support,
title = {Reliable support: Measuring calibration of likelihood ratios},
journal = {Forensic Science International},
volume = {230},
number = {1},
pages = {156-169},
note = {in {P}roc. EAFS 2012},
issn = {0379-0738},
doi = {https://doi.org/10.1016/j.forsciint.2013.04.014},
author = {Daniel Ramos and Joaquin Gonzalez-Rodriguez}
}

@article{wu2015,
    author = {Wu, Z. and Evans, N. and Kinnunen, T. and Yamagishi, J. and Alegre, F. and Li, H.},
    title = {Spoofing and Countermeasures for Speaker Verification: A Survey},
    journal = {Speech Communication},
    volume = {66},
    pages = {130-153},
    year = {2015},
}

@INPROCEEDINGS{11111888,
  author={Kurnaz, Oğuzhan and Kinnunen, Tomi H. and Hanilçi, Cemal},
  booktitle={in {P}roc. SIU 2025}, 
  title={Investigating the Potential of Multi-Stage Score Fusion in Spoofing-Aware Speaker Verification}, 
  year={2025},
  volume={},
  number={},
  pages={1-4},
  doi={10.1109/SIU66497.2025.11111888}}

@inproceedings{yakovlev24_interspeech,
  title     = {{Reshape Dimensions Network for Speaker Recognition}},
  author    = {Ivan Yakovlev and Rostislav Makarov and Andrei Balykin and Pavel Malov and Anton Okhotnikov and Nikita Torgashov},
  booktitle = {{Proc. Interspeech 2024}},
  pages     = {3235--3239},
  doi       = {10.21437/Interspeech.2024-2116},
  issn      = {2958-1796},
}

@inproceedings{Desplanques_2020,
  title     = {{ECAPA-TDNN}: Emphasized Channel Attention, Propagation and Aggregation in {TDNN} Based Speaker Verification},
  author    = {Brecht Desplanques and Jenthe Thienpondt and Kris Demuynck},
  booktitle = {Proc. Interspeech 2020},
  pages     = {3830--3834},
  doi       = {10.21437/Interspeech.2020-2650},
  issn      = {2958-1796},
}

@inproceedings{zhang22w_interspeech,
  title     = {Backend Ensemble for Speaker Verification and Spoofing Countermeasure},
  author    = {Li Zhang and Yue Li and Huan Zhao and Qing Wang and Lei Xie},
  booktitle = {Proc. Interspeech 2022},
  pages     = {4381--4385},
  doi       = {10.21437/Interspeech.2022-10259},
  issn      = {2958-1796}
}

@inproceedings{choi22b_interspeech,
  title     = {{HYU} Submission for the {SASV} Challenge 2022: Reforming Speaker Embeddings with Spoofing-Aware Conditioning},
  author    = {Jeong-Hwan Choi and Joon-Young Yang and Ye-Rin Jeoung and Joon-Hyuk Chang},
  booktitle = {Proc. Interspeech 2022},
  pages     = {2873--2877},
  doi       = {10.21437/Interspeech.2022-210},
  issn      = {2958-1796},
}

@inproceedings{zhang22f_interspeech,
  title     = {Norm-constrained Score-level Ensemble for Spoofing Aware Speaker Verification},
  author    = {Peng Zhang and Peng Hu and Xueliang Zhang},
  booktitle = {Proc. Interspeech 2022},
  pages     = {4371--4375},
  doi       = {10.21437/Interspeech.2022-470},
  issn      = {2958-1796},
}

@inproceedings{villalba24_asvspoof,
  title     = {The {SHADOW} team submission to the {ASV}spoof 2024 Challenge},
  author    = {Jesus Antonio Villalba and Tiantian Feng and Thomas Thebaud and Jihwan Lee and Shrikanth Narayanan and Najim Dehak},
  booktitle = {Proc. ASVspoof 2024},
  pages     = {36--42},
  doi       = {10.21437/ASVspoof.2024-6},
}

@article{Egozcue2003-ILR,
  author    = {Egozcue, J. J. and Pawlowsky-Glahn, V. and Mateu-Figueras, G. and Barceló-Vidal, C.},
  title     = {Isometric Logratio Transformations for Compositional Data Analysis},
  journal   = {Mathematical Geology},
  year      = {2003},
  volume    = {35},
  number    = {3},
  pages     = {279--300},
  doi       = {10.1023/A:1023818214614},
  publisher = {Springer}
}

@inproceedings{aliyev24_asvspoof,
  title     = {{INTEMA} system description for the {ASV}spoof5 Challenge: power weighted score fusion},
  author    = {Ali Aliyev and Alexander Kondratev},
  booktitle = {Proc. ASVspoof 2024},
  pages     = {152--157},
  doi       = {10.21437/ASVspoof.2024-22},
}

@inproceedings{duroselle24_asvspoof,
  title     = {Data augmentations for audio deepfake detection for the {ASV}spoof5 closed condition},
  author    = {Raphaël Duroselle and Olivier Boeffard and Adrien Courtois and Hubert Nourtel and Champion Pierre and Heiko Agnoli and Jean-François Bonastre},
  booktitle = {Proc. ASVspoof 2024},
  pages     = {16--23},
  doi       = {10.21437/ASVspoof.2024-3},
}

@inproceedings{rohdin24_asvspoof,
  title     = {{BUT} systems and analyses for the {ASV}spoof 5 Challenge},
  author    = {Johan Rohdin and et al.},
  booktitle = {Proc. ASVspoof 2024},
  pages     = {24--31},
  doi       = {10.21437/ASVspoof.2024-4},
}

@inproceedings{heo22_interspeech,
  title     = {Two Methods for Spoofing-Aware Speaker Verification: Multi-Layer Perceptron Score Fusion Model and Integrated Embedding Projector},
  author    = {Jungwoo Heo and Ju-Ho Kim and Hyun-seo Shin},
  booktitle = {Proc. Interspeech 2022},
  pages     = {2878--2882},
  doi       = {10.21437/Interspeech.2022-602},
  issn      = {2958-1796},
}

@inproceedings{lin22_interspeech,
  title     = {The {CLIPS} System for 2022 Spoofing-Aware Speaker Verification Challenge},
  author    = {Jucai Lin and Tingwei Chen and Jingbiao Huang and Ruidong Fang and Jun Yin and Yuanping Yin and Wei Shi and Weizhen Huang and Yapeng Mao},
  booktitle = {Proc. Interspeech 2022},
  pages     = {4367--4370},
  doi       = {10.21437/Interspeech.2022-320},
  issn      = {2958-1796},
}

@inproceedings{okhotnikov24_asvspoof,
  title     = {{IDV}oice team system description for {ASV}spoof5 Challenge},
  author    = {Anton Okhotnikov and et al.},
  booktitle = {Proc. ASVspoof 2024},
  pages     = {43--47},
  doi       = {10.21437/ASVspoof.2024-7},
}

@inproceedings{tran24_asvspoof,
  title     = {Parallel{C}hain {L}ab's anti-spoofing systems for {ASV}spoof 5},
  author    = {Thien Tran and Thanh Duc Bui and Panagiotis Simatis},
  booktitle = {Proc. ASVspoof 2024},
  pages     = {9--15},
  doi       = {10.21437/ASVspoof.2024-2},
}

@inproceedings{wang22ea_interspeech,
  title     = {The {DKU-OPPO} System for the 2022 Spoofing-Aware Speaker Verification Challenge},
  author    = {Xingming Wang and Xiaoyi Qin and Yikang Wang and Yunfei Xu and Ming Li},
  booktitle = {Proc. Interspeech 2022},
  pages     = {4396--4400},
  doi       = {10.21437/Interspeech.2022-11190},
  issn      = {2958-1796},
}

@inproceedings{martindonas24_asvspoof,
  title     = {{ASASVI}comtech: the {V}icomtech-{UGR} speech deepfake detection and {SASV} systems for the {ASV}spoof5 Challenge},
  author    = {Juan M. Martín-Doñas and Eros Rosello and Angel M. Gomez and Aitor Álvarez and Iván López-Espejo and Antonio M. Peinado},
  booktitle = {Proc. ASVspoof 2024},
  pages     = {144--151},
  doi       = {10.21437/ASVspoof.2024-21},
}

@inproceedings{chen24_asvspoof,
  title     = {{USTC-KXDIGIT} system description for {ASV}spoof5 Challenge},
  author    = {Yihao Chen and et al.},
  booktitle = {Proc. ASVspoof 2024},
  pages     = {109--115},
  doi       = {10.21437/ASVspoof.2024-16},
}

@ARTICLE{kurnaz2024optimizing,
  author={Kurnaz, Oğuzhan and Mishra, Jagabandhu and Kinnunen, Tomi H. and Hanilçi, Cemal},
  journal={IEEE Signal Processing Letters}, 
  title={Optimizing a-DCF for Spoofing-Robust Speaker Verification}, 
  year={2025},
  volume={32},
  number={},
  pages={1081-1085},
  doi={10.1109/LSP.2025.3545290}}

@inproceedings{Wang2024_ASVspoof5,
  title     = {{ASV}spoof 5: crowdsourced speech data, deepfakes, and adversarial attacks at scale},
  author    = {Xin Wang and et al.},
  booktitle = {Proc. The Automatic Speaker Verification Spoofing Countermeasures Workshop (ASVspoof 2024)},
  pages     = {1--8},
  doi       = {10.21437/ASVspoof.2024-1},
}

@inproceedings{jung2022aasist,
  title={{AASIST}: Audio anti-spoofing using integrated spectro-temporal graph attention networks},
  author={Jung, Jee-weon and Heo, Hee-Soo and Tak, Hemlata and Shim, Hye-jin and Chung, Joon Son and Lee, Bong-Jin and Yu, Ha-Jin and Evans, Nicholas},
  booktitle={Proc. ICASSP 2022},
  pages={6367--6371},
}

@inproceedings{jung2022sasv,
  author={Jee-weon Jung and Hemlata Tak and Hye-jin Shim and Hee-Soo Heo and Bong-Jin Lee and Soo-Whan Chung and Ha-Jin Yu and Nicholas Evans and Tomi Kinnunen},
  title={{SASV 2022: The First Spoofing-Aware Speaker Verification Challenge}},
  booktitle={Proc. Interspeech 2022},
  pages={2893--2897},
  doi={10.21437/Interspeech.2022-11270},
  issn={2958-1796}
}

@inproceedings{kingma2015adam,
  author       = {Kingma, Diederik P. and Ba, Jimmy},
  title        = {Adam: A Method for Stochastic Optimization},
  booktitle    = {Proc. ICLR 2015},
  editor       = {Bengio, Yoshua and LeCun, Yann}
}

@inproceedings{todisco2019asvspoof,
  title     = {{ASV}spoof 2019: Future Horizons in Spoofed and Fake Audio Detection},
  author    = {Massimiliano Todisco and et al.},
  booktitle = {Proc. Interspeech 2019},
  pages     = {1008--1012},
  doi       = {10.21437/Interspeech.2019-2249},
  issn      = {2958-1796},
}

@inproceedings{kurnaz24_asvspoof,
  title     = {Spoofing-robust speaker verification using parallel embedding fusion: {BTU} speech group's approach for {ASV}spoof5 Challenge},
  author    = {Oğuzhan Kurnaz and Selim Can Demirtaş and Aykut Büker Jagabandhu Mishra and Cemal Hanilçi},
  booktitle = {Proc. ASVspoof 2024},
  pages     = {138--143},
  doi       = {10.21437/ASVspoof.2024-20},
}

@inproceedings{Wu2022e,
  title     = {{Spoofing-Aware Speaker Verification by Multi-Level Fusion}},
  author    = {{Haibin Wu and Lingwei Meng and Jiawen Kang and Jinchao Li and Xu Li and Xixin Wu and Hung-yi Lee and Helen Meng}},
  booktitle = {{Proc. Interspeech 2022}},
  pages     = {{4357--4361}},
  doi       = {{10.21437/Interspeech.2022-920}},
  issn      = {{2958-1796}},
}

\end{document}